\documentclass[twocolumn]{aa}
\usepackage{graphicx}
\usepackage{txfonts}
\usepackage{comment}
\usepackage{float}
\usepackage[dvipsnames]{xcolor}
\usepackage{subfig}
\usepackage[utf8]{inputenc}
\usepackage[T1]{fontenc}
\usepackage{amsfonts}
\usepackage{longtable}
\usepackage{lscape}
\usepackage{nicefrac}
\usepackage[colorlinks=true,allcolors=blue]{hyperref}
\usepackage{mathrsfs}

\graphicspath{{./}{FIGS/}}

\begin{document} 
\title{Understanding the Multi-wavelength Thermal Dust Polarisation from the Orion Molecular Cloud in Light of the Radiative Torque Paradigm}
\author{Le Ngoc Tram\inst{1}\thanks{Corresponding author: Le Ngoc Tram \newline \email{nle@mpifr-bonn.mpg.de}},  
Thiem Hoang\inst{2,3},
Helmut Wiesemeyer\inst{1},
Isabelle Ristorcelli\inst{4},
Karl M. Menten\inst{1}, \\
Nguyen Bich Ngoc\inst{5,6},
Pham Ngoc Diep\inst{5,6}
}
\institute{$^{1}$ Max-Planck-Institut für Radioastronomie, Auf dem Hügel 69, 53121, Bonn, Germany\\
$^{2}$ Korea Astronomy and Space Science Institute, Daejeon 34055, Republic of Korea \\
$^{3}$ Korea University of Science and Technology, 217 Gajeong-ro, Yuseong-gu, Daejeon, 34113, Republic of Korea \\
$^{4}$ Univ. Toulouse, CNRS, IRAP, 9 Av. du colonel Roche, BP 44346, F-31028, Toulouse, France \\
$^{5}$ Department of Astrophysics, Vietnam National Space Center, Vietnam Academy of Science and Technology, 18 Hoang Quoc Viet, Hanoi, Vietnam \\
$^{6}$ Graduate University of Science and Technology, Vietnam Academy of Science and Technology, 18 Hoang Quoc Viet, Hanoi, Vietnam\\
} 
\date{accepted to A\&A}
\titlerunning{Examining Polarisation Spectrum in OMC-1}
\authorrunning{Tram et al., 2024}
\abstract
{Dust grains are important in various astrophysical processes and serve as indicators of interstellar medium structures, density, and mass. Understanding their physical properties and chemical composition is crucial in astrophysics. Dust polarisation is a valuable tool for studying these properties. The Radiative Torque (RAT) paradigm, which includes Radiative Torque Alignment (RAT-A) and Radiative Torque Disruption (RAT-D), is essential to interpret the dust polarisation data and constrain the fundamental properties of dust grains. However, it has been used primarily to interpret observations at a single wavelength. In this study, we analyse the thermal dust polarisation spectrum obtained from observations with SOFIA/HAWC+ and JCMT/POL-2 in the Orion Molecular Cloud 1 (OMC-1) region and compare the observational data with our numerical results using the RAT paradigm. In general, we show that the dense gas exhibits a positive spectral slope, whereas the warm regions show a negative one. We demonstrate that a one-layer dust (one-phase) model can only reproduce the observed spectra at certain locations and cannot match those with prominent V-shaped spectra (for which the degree of polarisation initially decreases with wavelength from 54 to $\sim$ 300$\,\mu$m and then increases at longer wavelengths). To address this, we improve our model by incorporating two dust components (warm and cold) along the line of sight, resulting in a two-phase model. This improved model successfully reproduces the V-shaped spectra. The best model corresponds to a mixture composition of silicate and carbonaceous grains in the cold medium. Finally, by assuming the plausible model of grain alignment, we infer the inclination angle of the magnetic fields in OMC-1. This approach represents an important step towards better understanding the physics of grain alignment and constraining 3D magnetic fields using dust polarisation spectra.} 

\maketitle
\section{Introduction}\label{sec:intro} 
The polarisation of light induced by interstellar dust provides valuable insight into the physical properties of dust, such as size, shape, composition, and alignment. 
The polarisation of starlight by interstellar dust was first observed in the 1940s (\citealt{1949Sci...109..166H,1949Sci...109..165H}), while the corresponding polarisation of thermal dust emission was detected in the 1980s (\citealt{1982MNRAS.200.1169C}). This observed polarisation of light requires the alignment of non-spherical grains with the magnetic field of the Milky Way (i.e. magnetic alignment). The widely accepted theory that describes the alignment of dust grains is Radiative Torque Alignment (RAT-A). This theory explains the magnetic alignment of irregular-shaped dust grains due to the effects of Larmor precession and RATs induced by their interaction with an incident anisotropic radiation field (see \citealt{1976Ap&SS..43..291D,1996ApJ...470..551D,2007MNRAS.378..910L}). RAT-A has been successful in explaining many polarimetric observations in diffuse gas and molecular clouds, as summarised in a review by \cite{2015ARA&A..53..501A}.

\begin{figure*}
    \centering
    \includegraphics[width=0.8\textwidth]{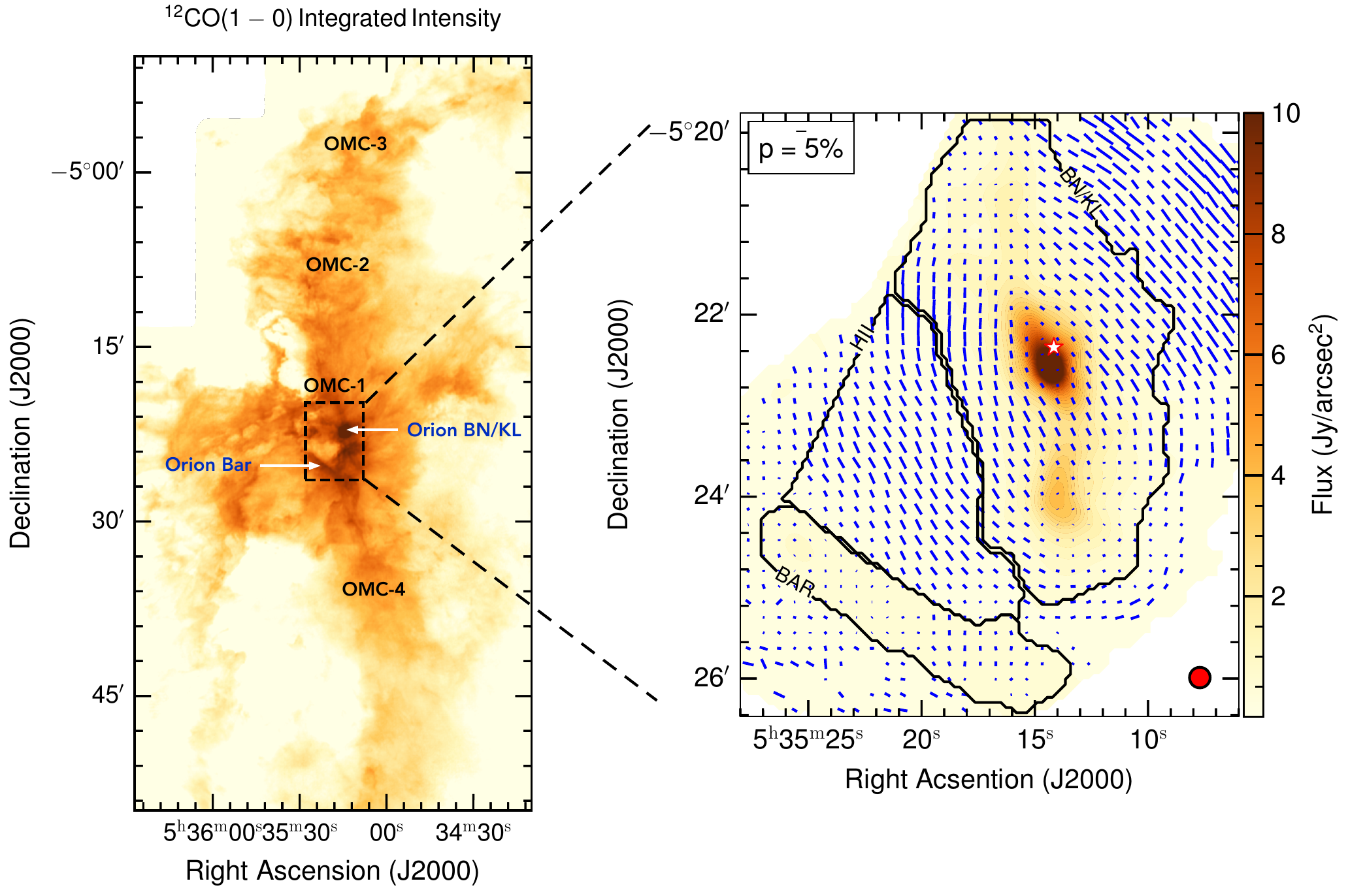}
    \caption{(\textbf{\textit{Left}}) The map showing the intensity of $^{12}$CO(1-0) integrated over the OMC A cloud, which is bounded by OMC-3 in the north and OMC-4 in the south. The focus region in this study is indicated by the black box. (\textbf{\textit{Right}}) The thermal dust polarisation was observed at a wavelength of 54$\,\mu$m using SOFIA/HAWC+. The background colour represents the total continuum intensity (Stokes-I). The blue segments indicate the orientation of the polarisation (E-vectors), with their length indicating the degree of polarisation. The black contour lines outline the Orion BN/KL, HII, and Bar. The red asterisk marks the location of the Orion BN/KL core. The CO data are obtained from the CARMA-NRO Orion Survey (\citealt{2018ApJS..236...25K}).}
    \label{fig:OMC-A}
\end{figure*}
Furthermore, \cite{2019NatAs...3..766H} and \cite{2019ApJ...876...13H} showed that when exposed to a strong radiation field, large dust grains acquire a remarkably high angular velocity and the largest dust grains in the population cannot survive intact because the centrifugal force within a grain exceeds the binding force that holds it together. This rotational disruption process is known as Radiative Torque Disruption (RAT-D). Extremely fast rotation of objects due to laser irradiation is demonstrated in laboratory experiments (e.g. \citealt{2018PhRvL.121c3602R,2018PhRvL.121c3603A}), and the disruptive effect of rotation is also seen in simulations (\citealt{2023arXiv230112889R}) and laboratory experiments (\citealt{2019EL....12745004H}). As the RAT-D mechanism sets an upper limit on the size distribution of dust grains, it has implications for various aspects of dust astrophysics, including light extinction, polarisation, and surface chemistry (see the review of \citealt{2020Galax...8...52H}). The combination of RAT-A theory and the RAT-D mechanism, referred to as the RAT paradigm, has been shown to reproduce a wider range of dust polarisation observations, extending to star-forming regions (see the review of \citealt{2022FrASS...9.3927T}).
 
However, most interpretations of polarisation data have been made based on data taken at a single wavelength or a combination of only a few wavelengths. To further validate the RAT theory, the next logical step is to synthesise the observed multi-wavelength polarisation, also known as the polarisation spectrum predicted by the RAT theory, and confront it with observational data. Previous observational analyses (see, e.g. \citealt{2016ApJ...824...84G,2019ApJ...872..197S} and references therein) have shown that the slope of the polarisation spectrum can vary locally within an interstellar cloud. In a recent study, \cite{2021ApJ...907...46M} used observations made in four bands of the High Resolution Airborne Wideband Camera Plus (HAWC+, \citealt{2018JAI.....740008H}) on board the Stratospheric Observatory for Infrared Astronomy (SOFIA) centred at 54, 89, 154, and 214$\,\mu$m towards Orion Molecular Cloud 1 (OMC-1) in Orion A cloud. They found that the slope decreases in the cooler regions and increases in the warmer regions. 

\cite{2022MNRAS.512.1985F} attempted to interpret the observed data in the NGC 2071 star-forming region using the modelling approach introduced in \cite{2018A&A...610A..16G}. However, they found that the models were unable to accurately reproduce the observations, even when incorporating parameter variations. Other studies on multiple dust polarisation observations in various clouds have reported a 'V-shape' in the spectra (e.g. \citealt{1999ApJ...516..834H,2008ApJ...679L..25V,2012ApJS..201...13V}), where the polarisation degree initially decreases with wavelength and then increases with longer wavelengths. To understand these observations, \cite{2000PASP..112.1215H, 2023arXiv231017211S} have suggested the need to consider multiple dust components with different temperatures and alignment efficiencies along the line of sight. In the latter work, the authors emphasized that grain composition plays a significant role in determining the V-shape.

In this study, we revisit the polarisation spectrum in OMC-1 as shown in Figure \ref{fig:OMC-A}, which hosts dense molecular material (the Kleinmann-Low nebula) and a dense photodissociation region (the Orion bar). We refer to \cite{GenzelStutzki1989} for a complete overview. This well-studied environment provides an excellent opportunity to study the physics of magnetically aligned grains under various physical conditions. Our objective is to quantitatively compare the predicted polarisation spectrum based on the RAT paradigm with the observed data from the far-infrared (53, 89, 154, and 214$\,\mu$m) reported in \cite{2021ApJ...907...46M} to submillimetre wavelengths (450 and 850$\,\mu$m) reported in \cite{2021ApJ...913...85H}.

The structure of this paper is as follows. Section \ref{sec:obs} presents our analysis of the multi-wavelength thermal dust polarisation observations. We discuss the interpretation of the observed spectra using the RAT paradigm in Section \ref{sec:model}. Finally, our discussions and conclusions are provided in Sections \ref{sec:Discussion} and \ref{sec:Conclusion}, respectively.

\section{Observed polarisation Spectrum}\label{sec:obs}
\subsection{Ancillary Data}
In this study, we utilized the thermal dust polarisation data obtained from the HAWC+ camera onboard the SOFIA and the POL-2 polarimeter installed on the James Clerk Maxwell Telescope (JCMT). The SOFIA/HAWC + observations at wavelengths 54, 89, 154, and 214$\,\mu$m were previously reported in \cite{2019ApJ...872..187C}, while the JCMT/Pol-2 observations at 450 and 850$\,\mu$m were published in \cite{2021ApJ...913...85H}. To ensure consistency, we applied a smoothing process using convolution kernels to all Stokes I, Q and U maps with the largest full width at half maximum (FWHM) value of $18\rlap{.}''2$. Subsequently, we re-gridded all the smoothed images to a uniform pixel size of $4\rlap{.}''55$. The degree, angle and respective errors of the debiased polarisation were estimated following Appendix \ref{app:obs_syn}. Spectra were extracted from spatial regions for which data were available for all wavelength bands. In addition, we added the data obtained during the APEX/PolKA commissioning observations at 870$\,\mu$m (\citealt{2014PASP..126.1027W}) to the spectra for comparison only and excluded these data points from the fitting process in this study.

\subsection{polarisation Spectra}
At each wavelength, we first computed the average degree of polarisation ($p(\%)$) at each pixel within a two-beam kernel and then created the $p(\%)$ maps and the associated error before stacking them into a data cube. The spatial area of this data cube is defined by the condition that all six wavelengths must fulfill the circular beam, ensuring that the different data sets cover the same area of interest. Furthermore, the value of $p(\%)$ in the local pixel is set to null if the circular beam surface is filled less than 25\% by the observed data. 

\begin{figure}
    \centering
    \includegraphics[width=0.5\textwidth]{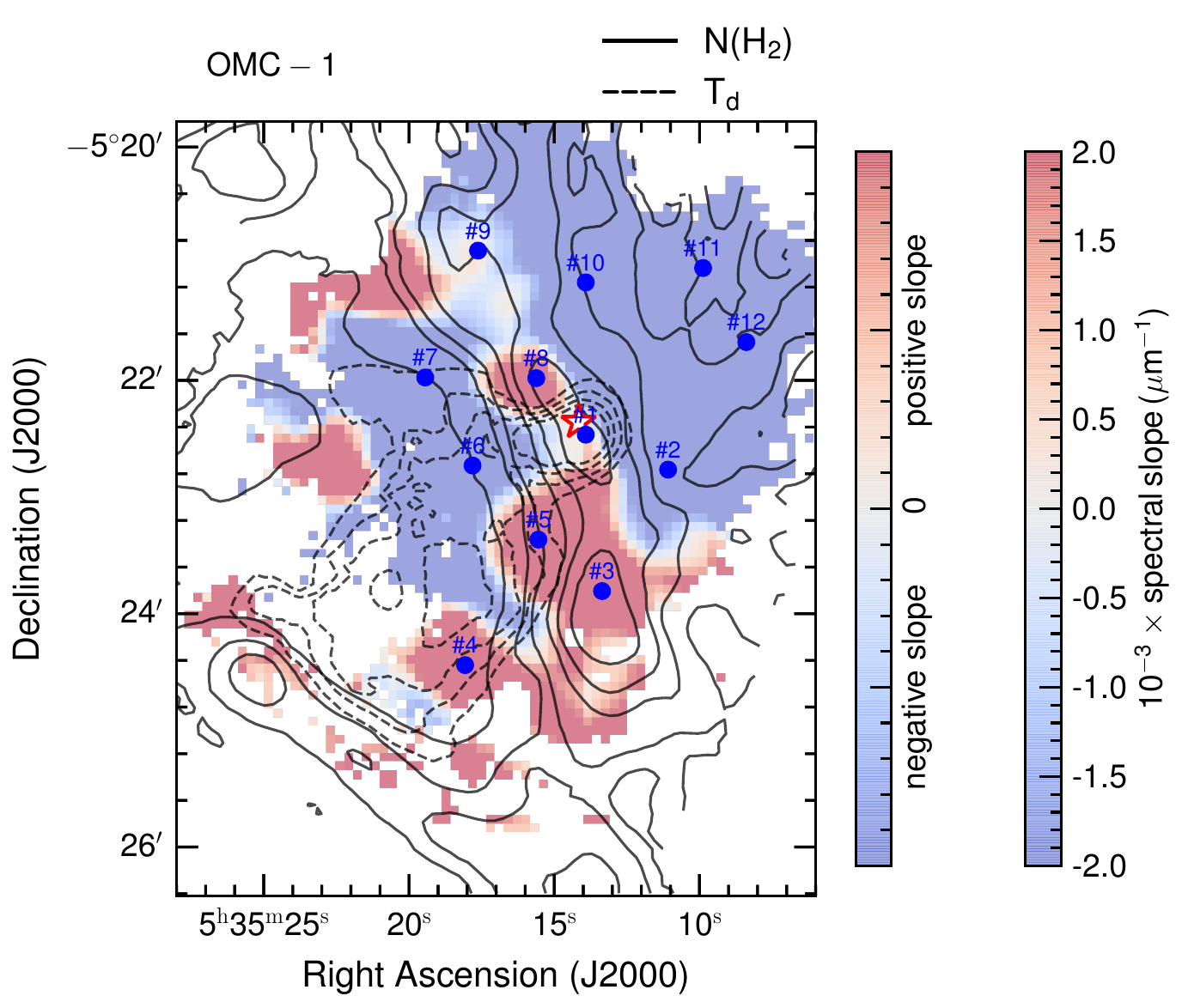}
    \caption{The spatial distribution of the spectral slope of the polarisation spectrum in OMC-1. The reddish region shows a rising polarisation spectrum, whereas the blueish region shows a spectrum with a falling slope. This map is overlayed with dust temperature (solid black lines with $T_{\rm d}\geq 60\,$K with a step of 5$\,$K) and gas density (dashed black line with $N(\rm H_{2}) = 1,\, 2,\,3,\,6,\,10,\,20 \times 10^{22}\,\rm cm^{-2}$ or $A_{\rm V}\geq 12.4\,$magnitude). The blue dots indicate the 12 positions where the spectra are used to compare with numerical models.} \label{fig:pol_spec_obs}
\end{figure}
To know whether the polarisation spectrum, in general, increases or falls, we used a linear line to fit these spectra from 54 to 850$\,\mu$m in each pixel, similar to that used by \cite{2021ApJ...907...46M}, to determine the slope sign of the polarisation spectrum. The resulting slope map can be seen in Figure \ref{fig:pol_spec_obs}. A positive (negative) value of the slope indicates an increase (decrease) in the polarisation spectrum. We observe a rising spectrum (indicated by reddish colours) in the main filament structure of Orion and the Bar, while a falling spectrum (indicated by blueish colours) is observed elsewhere. In correlation with dust temperature and gas column density derived from the continuum dust SED fitting \footnote{The maps of dust temperature and gas column density were obtained from the SED fitting conducted in \cite{2019ApJ...872..187C}.}, it is observed that the increasing spectrum is closely correlated with dense gas, where dust temperature is not the hottest, except at the Orion BN/KL core.

In order to qualitatively compare observations and modelling, we used the average spectra over two beam sizes at 12 locations distributed throughout the OMC-1 region, as indicated by the blue dots in Figure \ref{fig:pol_spec_obs}. The resulting spectra and the corresponding slopes are shown in Figure \ref{fig:modelvsobs_1phase}, including error bars that represent the statistical error associated with the average kernel. It is evident that the shape and slope of the spectrum vary across the cloud, with some locations exhibiting a pronounced V-shaped shape.

\begin{table}[!ht]
\centering
\caption{The main input parameters of the modelling (radiation field and gas density) and the output parameters (maximum alignment efficiency cross the inclination angle of magnetic field and the size-distribution power index) from the best-fitting one-phase model. Besides, the maximum grain size is fixed as $2\,\mu$m, the mean wavelength of the field is fixed as $0.5\,\mu$m, and tensile strength is fixed as $10^{7}\,\rm erg\,cm^{-3}$.} 
\label{tab:one-phase_model}     
\renewcommand{\footnoterule}{}  
\begin{tabular}{c c c c c c}     
\multicolumn{6}{c}{One-phase Model} \\
\hline\hline       
Regions & $U$ & $n_{\rm H}$ & $f_{\rm max}\times \sin^{2}\psi$ & $\beta$ & $\chi^{2}$ \\
{} & {} &  $\rm (cm^{-3})$ & & & \\
\hline
\#1     & 12099& $9.05\times 10^{5}$ & 0.80 & -4.81 & 0.72 \\
\#2     & 555  & $1.89\times 10^{5}$ & 0.33 & -3.00 & 17.88 \\
\#3     & 372  & $1.09\times 10^{6}$ & 0.08 & -3.66 & 6.60  \\
\#4     & 5117 & $3.09\times 10^{4}$ & 0.20 & -3.14 & 4.17  \\
\#5     & 3153 & $1.63\times 10^{5}$ & 0.31 & -3.93 & 6.68  \\
\#6     & 7228 & $5.16\times 10^{4}$ & 1.00 & -3.85 & 6.88  \\
\#7     & 2403 & $5.40\times 10^{4}$ & 0.75 & -3.00 & 12.52 \\
\#8     & 857  & $7.62\times 10^{5}$ & 1.00 & -4.55 & 2.98  \\
\#9     & 52   & $4.98\times 10^{5}$ & 0.13 & -3.00 & 12.33 \\
\#10    & 184  & $2.54\times 10^{5}$ & 0.29 & -3.00 & 46.76 \\
\#11    & 59   & $1.58\times 10^{5}$ & 0.42 & -3.00 & 116.37 \\
\#12    & 45   & $1.39\times 10^{5}$ & 0.45 & -3.00 & 317.91 \\
\hline
\hline   
\end{tabular}

\vspace{5mm}
\caption{Similar to Table \ref{tab:one-phase_model} but for the two-phase model. The total and polarized emission in the second phase are scaled by $f_{\rm scale}$ and $f_{\rm heat}$ relatively to the first phase with $f_{\rm heat}=T^{\rm first-phase}_{\rm dust}/T^{\rm second-phase}_{\rm dust}$.} 
\label{tab:two-phase_model}     
\renewcommand{\footnoterule}{} 
\begin{tabular}{c c c c c c}     
\multicolumn{6}{c}{Two-phase Model} \\
\hline\hline        
Regions & $f_{\rm max}\times \sin^{2}\psi$ & $\beta$ & $f_{\rm heat}$ & $f_{\rm scale}$ & $\chi^{2}$ \\
\hline
\#1     & 0.53 & -4.50 & 2.14 & 6.85  & 0.56 \\
\#2     & 0.29 & -3.53 & 3.06 & 39.89 & 10.73\\
\#3     & 0.57 & -4.00 & 2.13 & 181.46& 3.41 \\
\#4     & 0.58 & -4.23 & 1.53 & 60.42 & 2.15 \\\
\#5     & 0.69 & -4.37 & 1.24 & 104.03& 2.61 \\
\#6     & 0.94 & -4.22 & 5.37 & 39.67 & 5.29 \\
\#7     & 0.94 & -4.17 & 3.42 & 31.25 & 3.74 \\
\#8     & 0.58 & -4.20 & 1.70 & 11.42 & 2.53 \\
\#9     & 0.16 & -3.00 & 1.84 & 161.87& 2.30 \\
\#10    & 0.32 & -3.37 & 2.50 & 113.48& 16.33\\
\#11    & 0.34 & -3.00 & 2.35 & 36.86 & 18.00\\
\#12    & 0.38 & -3.02 & 2.31 & 40.52 & 26.19\\

\hline
\hline   
\end{tabular}
\end{table}

\section{Modelled polarisation Spectrum}\label{sec:model}
\subsection{One-phase Dust Model}
We utilize a simplified model based on the RAT paradigm, as described in \cite{2020ApJ...896...44L} and updated in \cite{2021ApJ...906..115T}\footnote{The \textsc{python}-based source code is publicly available at \href{https://github.com/lengoctram/DustPOL-py}{https://github.com/lengoctram/DustPOL-py}}. The model incorporates various input parameters that characterize local physical conditions (such as radiation field and gas density), grain properties (including size and shape), grain composition (such as silicate, carbonaceous, and mixtures), and grain compactness (tensile strength). It is worth noticing that our model accounts for neither depolarisation caused by magnetic field fluctuation along the line of sight nor for the radiative transfer effects. In this work, we take into account the inclination effect of the magnetic field by introducing a new parameter, which is the angle of inclination $\psi$ of the regular magnetic field to the sight line. This inclined magnetic field reduces the degree of net polarisation by a factor of $\sin^{2}\psi$. The fundamentals of our model are described in Appendix \ref{app:model_onephase}. 

Figure \ref{fig:one-phase-model} visualises the polarisation spectrum predicted by our one-phase model. In the case that only silicate grains are aligned, the degree of polarisation first increases and then decreases towards a longer wavelength. Ideal carbonaceous grains are diamagnetic materials; therefore, they cannot be coupled and aligned efficiently with magnetic fields as silicate grains (we refer to \citealt{2023ApJ...954..216H} for a study of the alignment of carbonaceous grains). However, if the carbon grains adhere to the silicate grains (referred to as the mixture in this paper), they can be 'passively aligned' with the magnetic field and then produce a polarisation signal. In this case, the degree of polarisation is higher in amplitude compared to that in the previous case, and the spectrum gradually increases at a longer wavelength. One can see that the higher radiation field makes the spectrum rise at a shorter wavelength, and the degree is higher if the disruption effect is neglected (left panel) compared to the lower radiation field medium. 

When the disruption effect is considered (right panel), the depletion of the largest grains for a sufficiently intense radiation field causes the polarisation degree to decrease, which can be even smaller than in the case of a low radiation field. Note that the value of $p(\%)$ depends on the input parameters, and the value shown in Figure \ref{fig:one-phase-model} is subject to variability for different sets of parameters.
 
\begin{figure*}[!ht]
    \centering
    \includegraphics[width=0.45\textwidth]{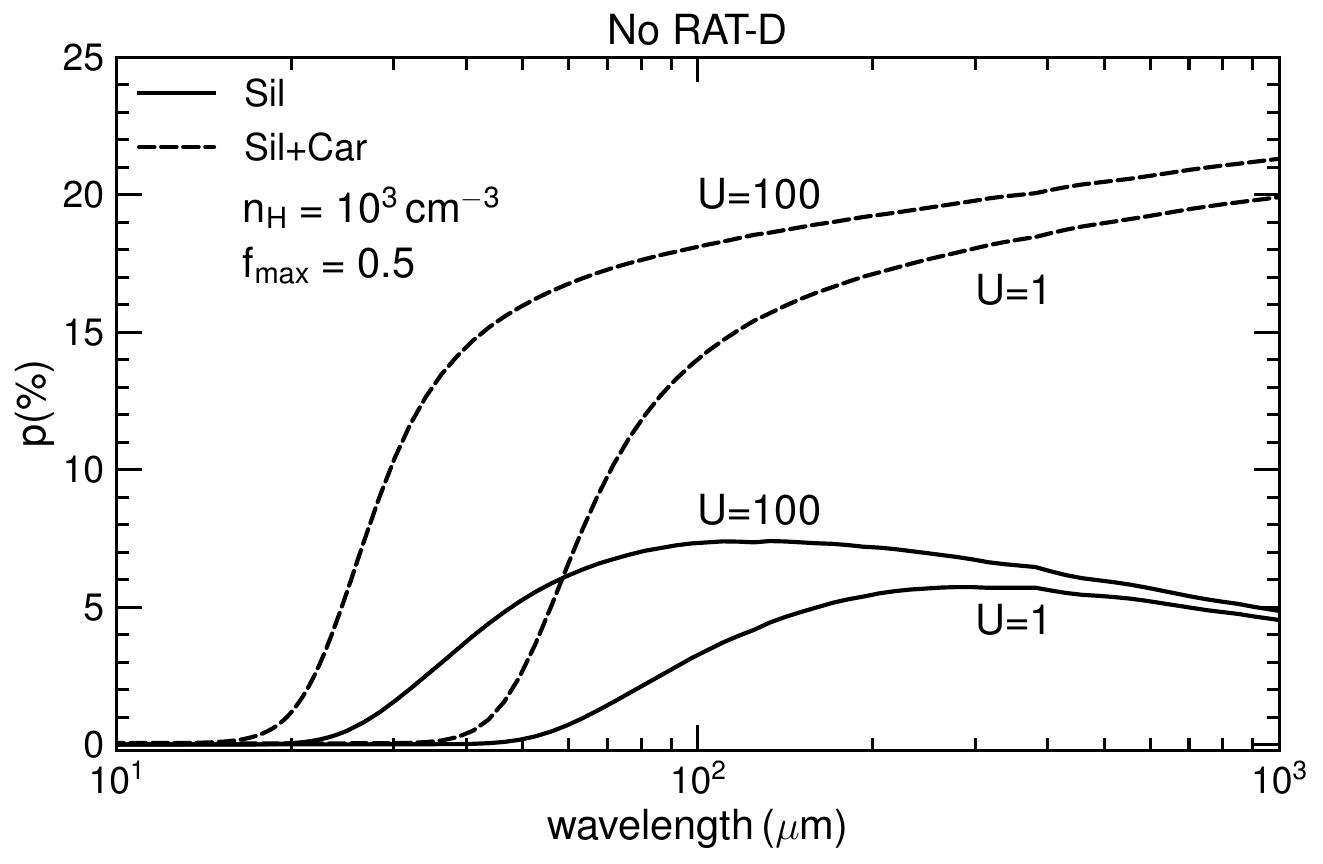}
    \includegraphics[width=0.45\textwidth]{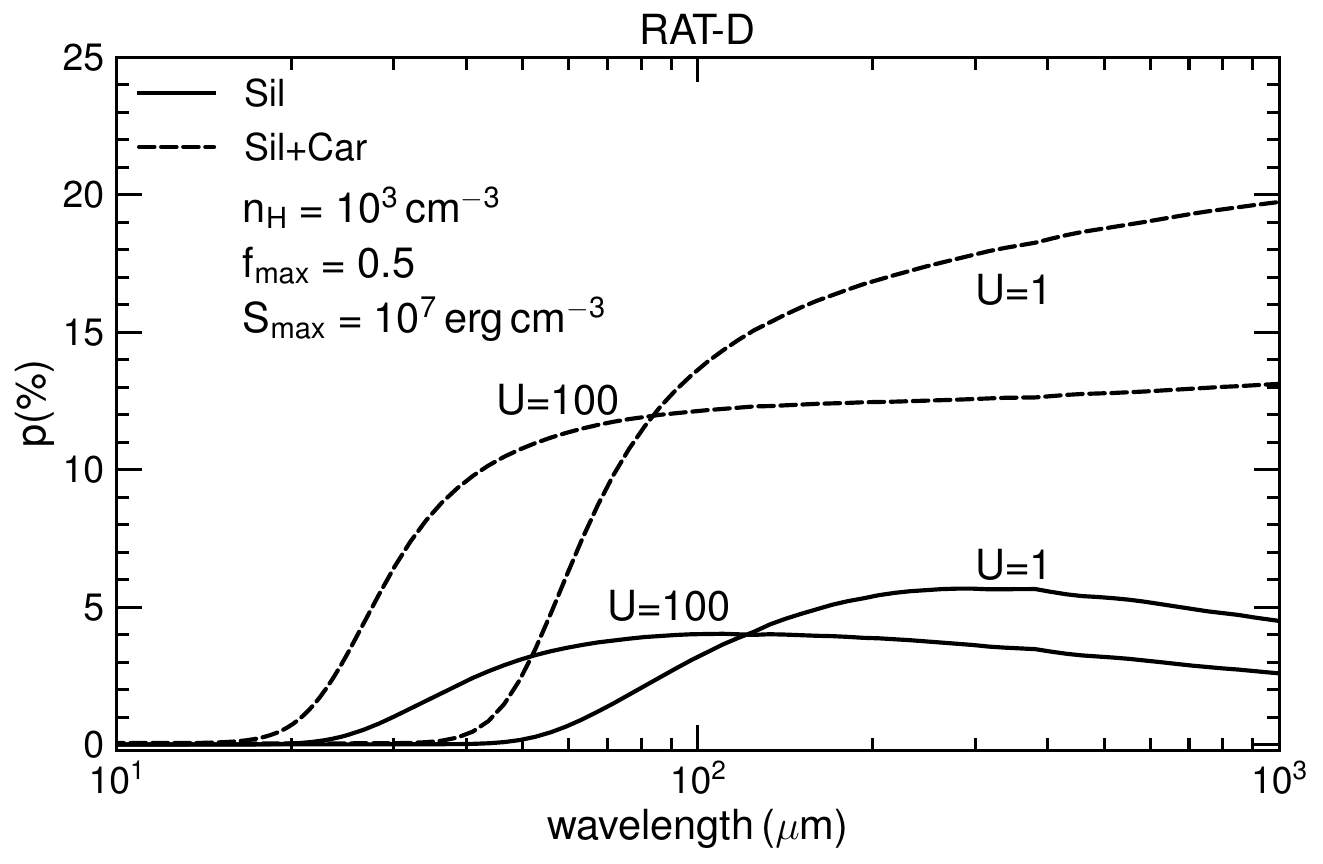}
    \caption{The polarisation spectra predicted by our one-phase modelling for two radiation strengths of $U=1$ (i.e. the standard ISM radiation field) and $U=100$ (i.e. a hundred times higher than the standard ISM). Solid lines show the spectra for only aligned silicates, while dashed lines show the aligned silicate and carbonaceous mixture. The disruption by RAT-D is considered in the right panel for the composite grains with $S_{\rm max}=10^{7}\,\rm erg\,cm^{-3}$. The total intensity is a summation of both silicate and carbonaceous grains.}
    \label{fig:one-phase-model}
    \includegraphics[width=0.45\textwidth]{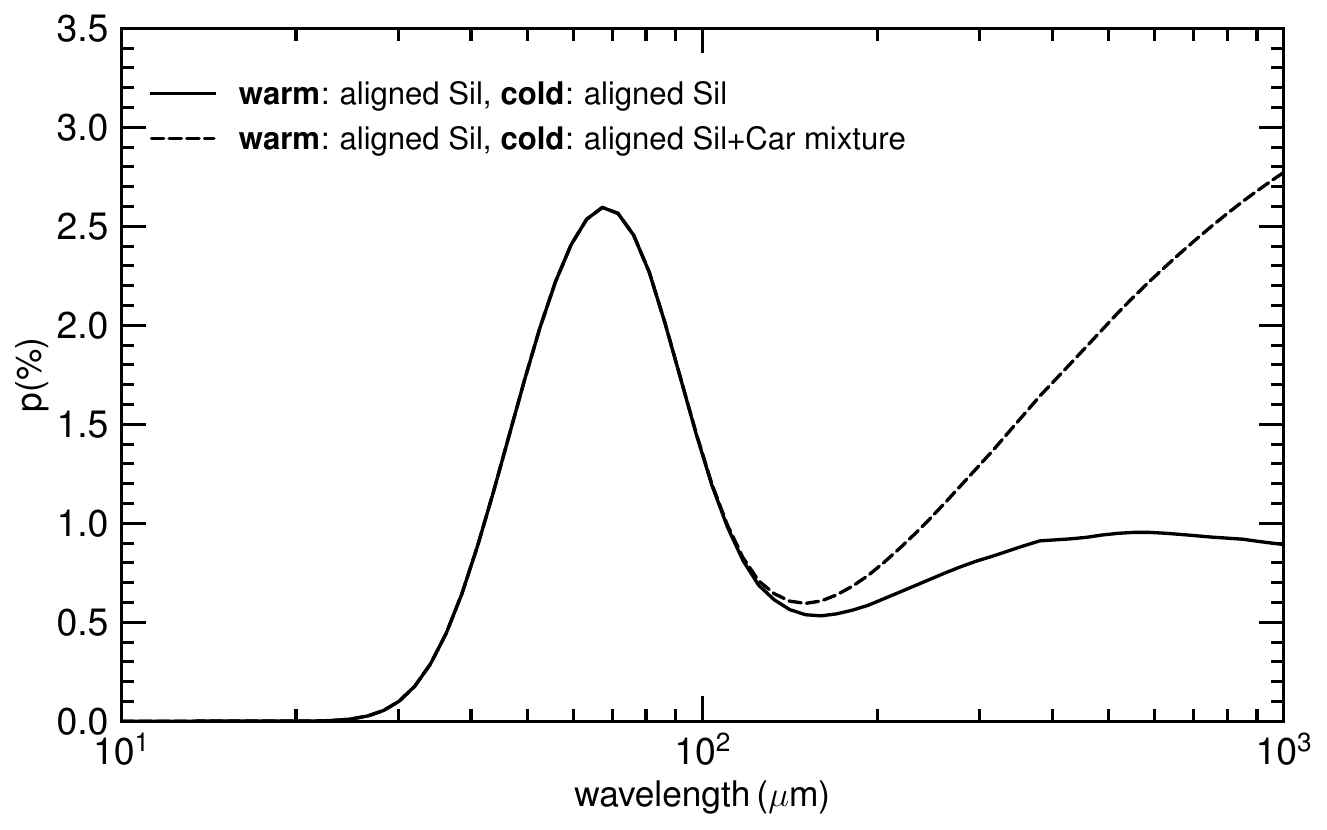}
    \includegraphics[width=0.45\textwidth]{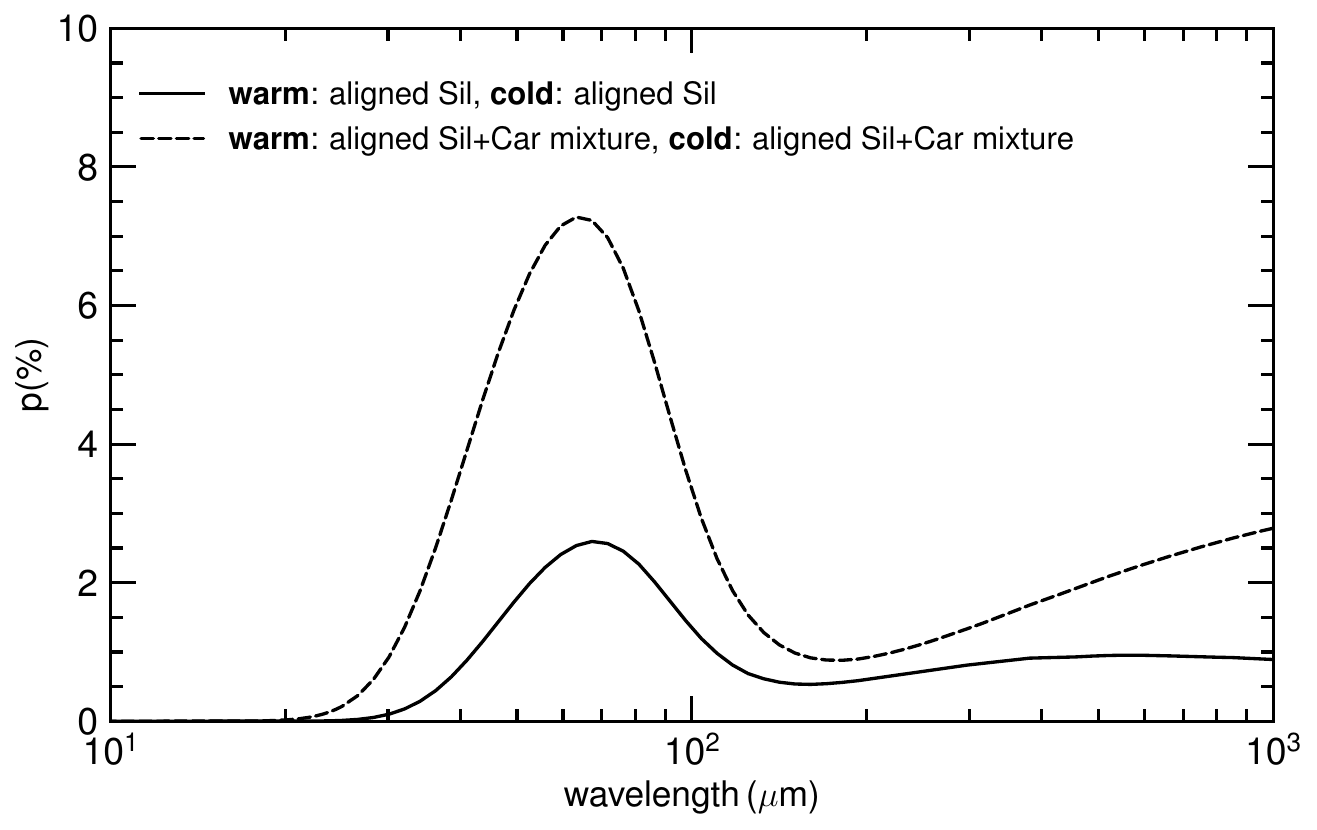}
    \caption{The polarisation spectrum predicted by our two-phase model. On the left, only aligned silicate produces polarisation in the first phase (warm phase), while either aligned silicate (solid line) or aligned mixed silicate and carbonaceous (dashed line) in the second phase (cold phase) produce polarisation. The right panel shows the case of mixed composition in the warm phase. This mixture increases $p(\%)$ in the warm phase and consequently shallows the V-shaped compared to that of only silicate showing in the left panel. We adopted $f_{\rm max}=0.5$, $f_{\rm heat}=5$ with $U=40$ and $n_{\rm H}=10^{3}\,\rm cm^{-3}$.}
    \label{fig:two-phase-model}
\end{figure*}

We used local values of dust temperature ($T_{\rm d}$) and gas column density ($N(\rm H_{2})$) that were fit to the spectral energy distribution (SED) of the dust (see Figure 2 in \citealt{2019ApJ...872..187C}). Subsequently, we determined the local radiation intensity as $U\simeq (T_{\rm d}/16.4\, \rm K)^{6}$, where $U=u_{\rm rad}/(8.64\times 10^{-13}\,\rm erg\, cm^{-3})$ represents the ratio of the radiation density at the specific location to the radiation density in the surrounding area (\citealt{2011piim.book.....D}). We estimate the density of the local gas volume as $n_{\rm H} = 2\times N(\rm H_{2})/W$, where $W$ denotes the depth of the cloud. For OMC-1, $W\simeq 0.15\,$pc (see \citealt{2017ApJ...846..122P,2019ApJ...872..187C}), which is in agreement with the turbulent correlation length estimated in \cite{2021ApJ...908...98G}. In our model, two main parameters are varied: the product of the maximum alignment efficiency and the depolarisation by magnetic field fluctuation ($f_{\rm max}\times \sin^{2}\psi$), and the power index of the grain size distribution ($\beta$). The adopted parameters can be found in Table \ref{tab:one-phase_model}.

We used LMFIT least squares minimisation (\citealt{2016ascl.soft06014N}), which extends the Levenberg–Marquardt method to restrict the best models, which were then used to fit the observed spectra. These fitted spectra are represented by the solid lines in Figure \ref{fig:modelvsobs_1phase}. The accuracy of the fitting is demonstrated in Appendix \ref{app:model_fitting}. Our model is capable of reasonably replicating the observed spectra in certain regions, such as \#1, \#3, \#4, \#5, \#6 and \#8. However, our model does not fit the spectra well in other areas, which leads to unconstrained parameters at those locations. Interestingly, the spectra that our model fails to reproduce exhibit a distinct 'V-shape' pattern.

\begin{figure*}
    \centering
    \includegraphics[width=0.95\textwidth]{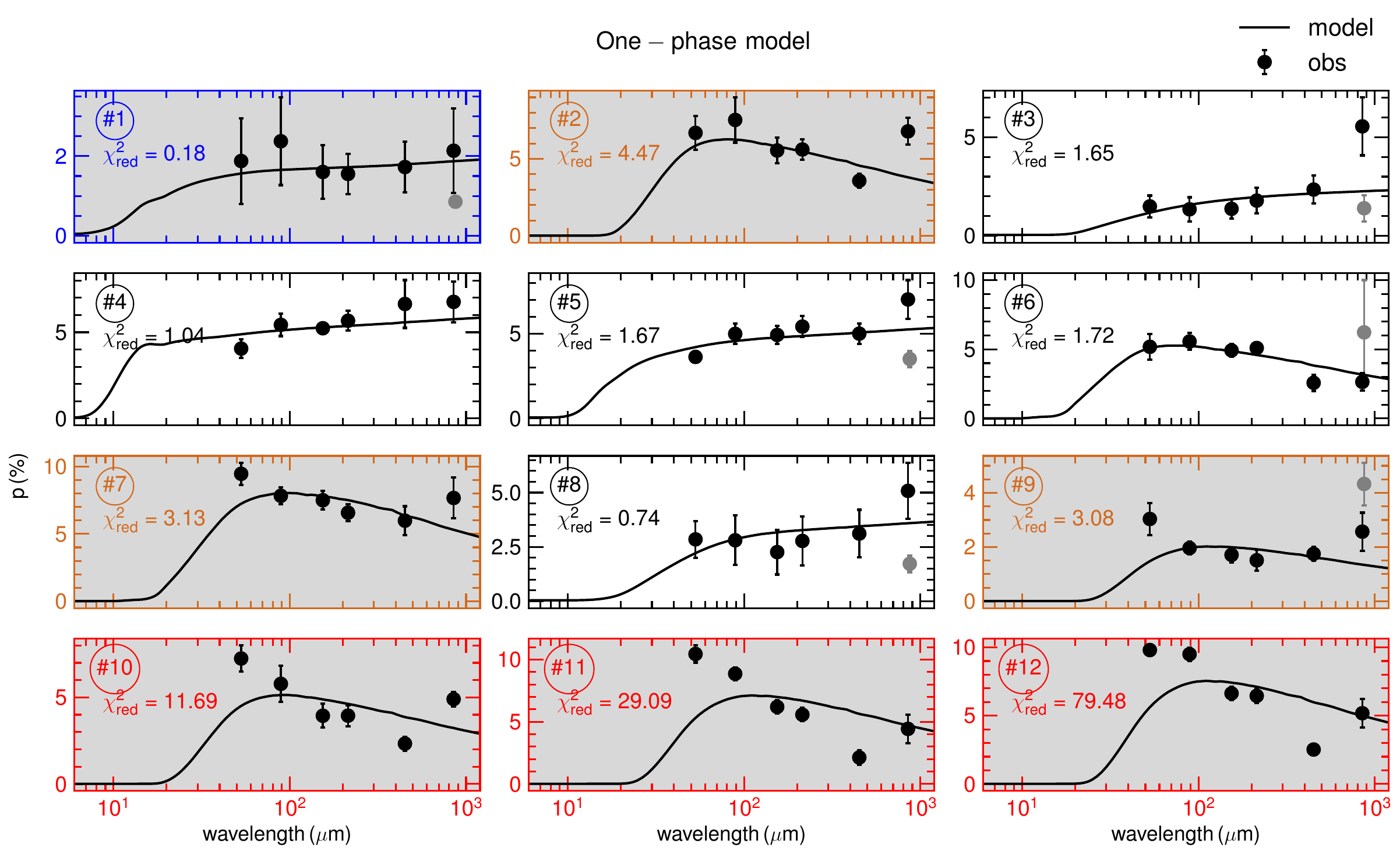}
    \caption{The polarisation spectrum observed by SOFIA/HAWC+ and JCMT/Pol-2 (black dots) at 12 different locations in OMC-1 as shown in Figure \ref{fig:pol_spec_obs} with the best one-phase model. The reduced chi-square ($\chi^{2}_{\rm red}$) indicates the 'good goodness' of the fit. The standard $\chi^{2}$ is shown in Table \ref{tab:one-phase_model}. For comparison, we show the APEX/PolKA observations at 870$\,\mu$m as a gray dot, but exclude them from the fitting. For visual guidance, we colour the frame blue for $\chi^{2}_{\rm red}<0.5$, brown for $3<\chi^{2}_{\rm red}<10$, and red for $\chi^{2}_{\rm red}\geq10$.}
    \label{fig:modelvsobs_1phase}
    \includegraphics[width=0.95\textwidth]{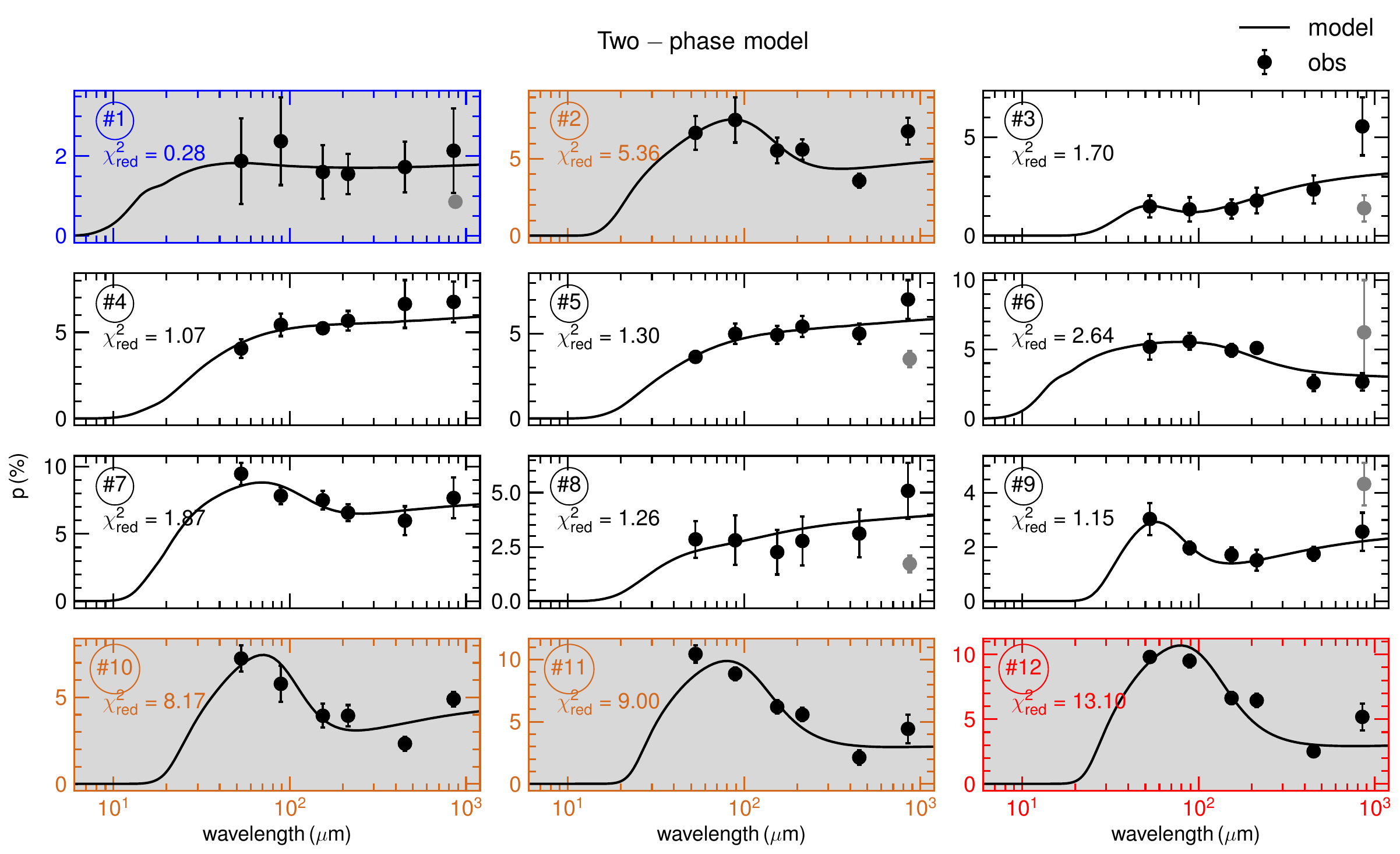}
    \caption{Similar to Figure \ref{fig:modelvsobs_1phase} but with the best two-phase model. Compared to the one-phase model, the two-phase model significantly improves the quality of the fits, as illustrative by $\chi^{2}_{\rm red}$ or standard $\chi^{2}$ in Table \ref{tab:two-phase_model}. The presence of a cold dust layer is insignificant in \#1, \#4, \#5, and \#8, while the model cannot reproduce the V-shaped spectrum with a sharply rising slope in \#2, \#3, and \#12.}
    \label{fig:modelvsobs_2phase}
\end{figure*}

\subsection{Two-phase Dust Model}
As previously mentioned, the polarisation spectrum exhibits a 'V-shape' pattern, which cannot be accurately reproduced by the one-phase model. This phenomenon has been commonly observed in the literature (see, e.g. \citealt{1999ApJ...516..834H,2008ApJ...679L..25V,2012ApJS..201...13V}), and it is believed that multiple dust components are required along the line of sight \citep{2000PASP..112.1215H,2023arXiv231017211S}. 
In the case of OMC-1, a 3D dust map of OMC-A reconstructed by \cite{2022ApJ...930L..22R} has revealed the presence of multiple dust layers. Therefore, we have updated our one-phase model to a two-phase model, consisting of one warm layer and one cold layer, following the procedure outlined in \cite{2023arXiv231017211S}. The governing equations for estimating the total and polarised intensities are essentially the same as those for the first layer. However, the relative contribution of the two phases to these intensities is parameterized by $f_{\rm scale}$, and the relative difference in dust temperature (radiation field) is parameterized by $f_{\rm heat}$. We have made the critical assumption that the grain size distribution is identical in both phases and therefore we have varied four parameters ($f_{\rm max}\times \sin^{2}\psi$, $\beta$, $f_{\rm scale}$, and $f_{\rm heat}$). The details of the two-phase model are described in Appendix \ref{app:model_twophase}. 

Figure \ref{fig:two-phase-model} shows the predictions of the two-phase model. In general, this model can reproduce the observed V-shaped polarisation spectrum. Moreover, a mixture of silicate and carbonaceous in the second phase results in a steep continuous increase of $p(\%)$ towards longer wavelengths (longer than 1$\,$mm), thus making the spectrum's V-shaped more pronounceable. In contrast, the model with only aligned silicate grains causes $p(\%)$ to increase before decreasing in the millimetre range. It should be noted that, in the first phase, the shape of the spectrum is invariant, whether the alignment of only silicate or mixture to carbonaceous is considered, yet $p(\%)$ is higher for a mixture of dust composition. 

Figure \ref{fig:modelvsobs_2phase} shows the best-fit representations of the observations achieved by utilising the two-phase model. The convergence of the fit is shown in Appendix \ref{app:model_fitting}. The most optimal values for the parameters of this model can be found in Table \ref{tab:two-phase_model}. Compared with the one-phase model, the fit quality is generally improved from a statistical perspective, and the two-phase model successfully reproduces the V-shaped pattern observed in the spectra. We note that the shape of the polarisation spectra is similar to the one-phase model in regions \#1, \#4, \#5, and \#8, and the model encounters difficulties in accurately reproducing the V-shaped spectrum with a steeply increasing slope in regions \#2, \#3, and \#12.

\section{Discussions}\label{sec:Discussion}
\subsection{Variation of Magnetic Fields at Different Wavelengths} \label{sec:field_tangling}
\begin{figure*}
    \centering
    \includegraphics[width=0.33\textwidth]{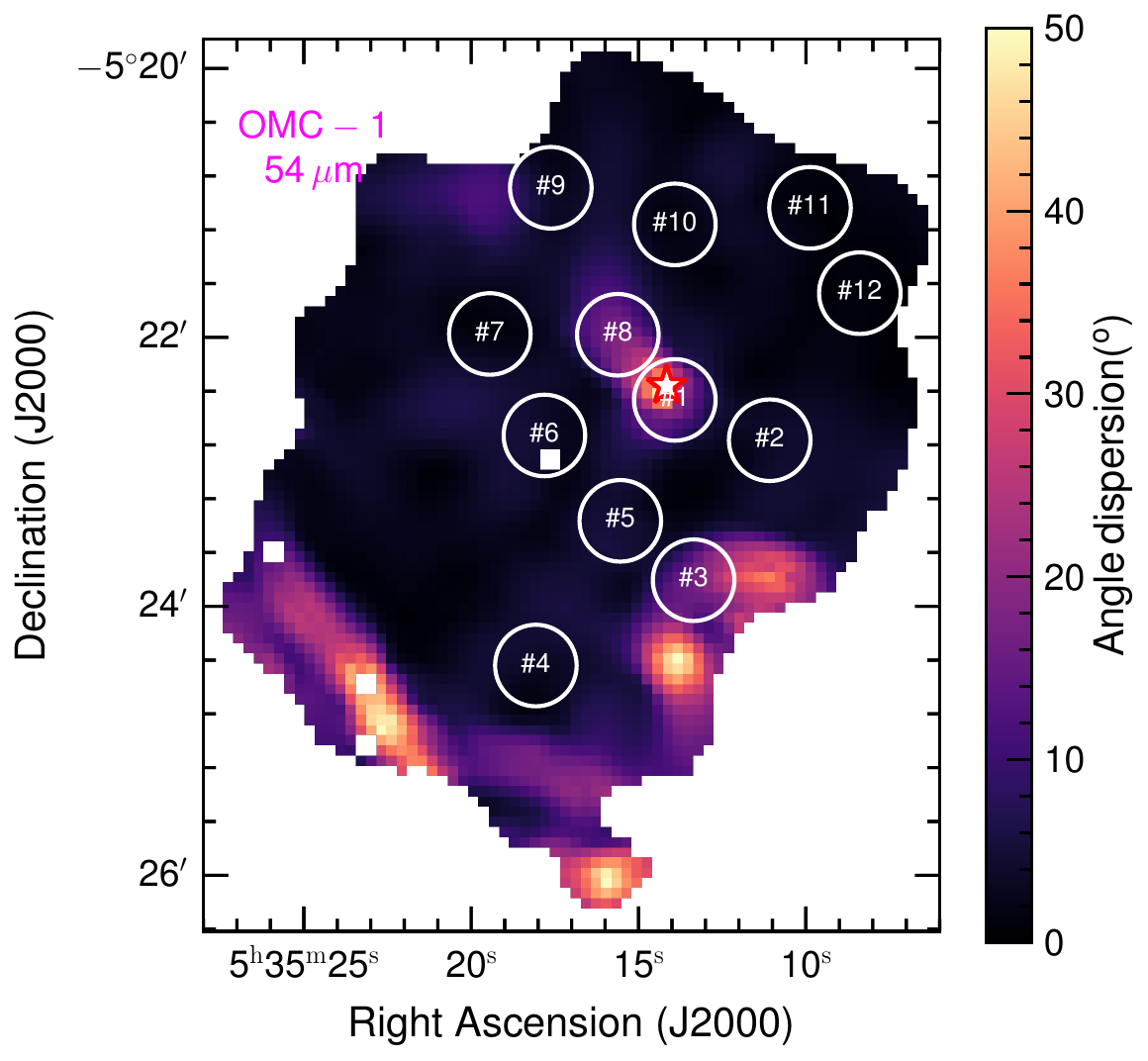}
    \includegraphics[width=0.33\textwidth]{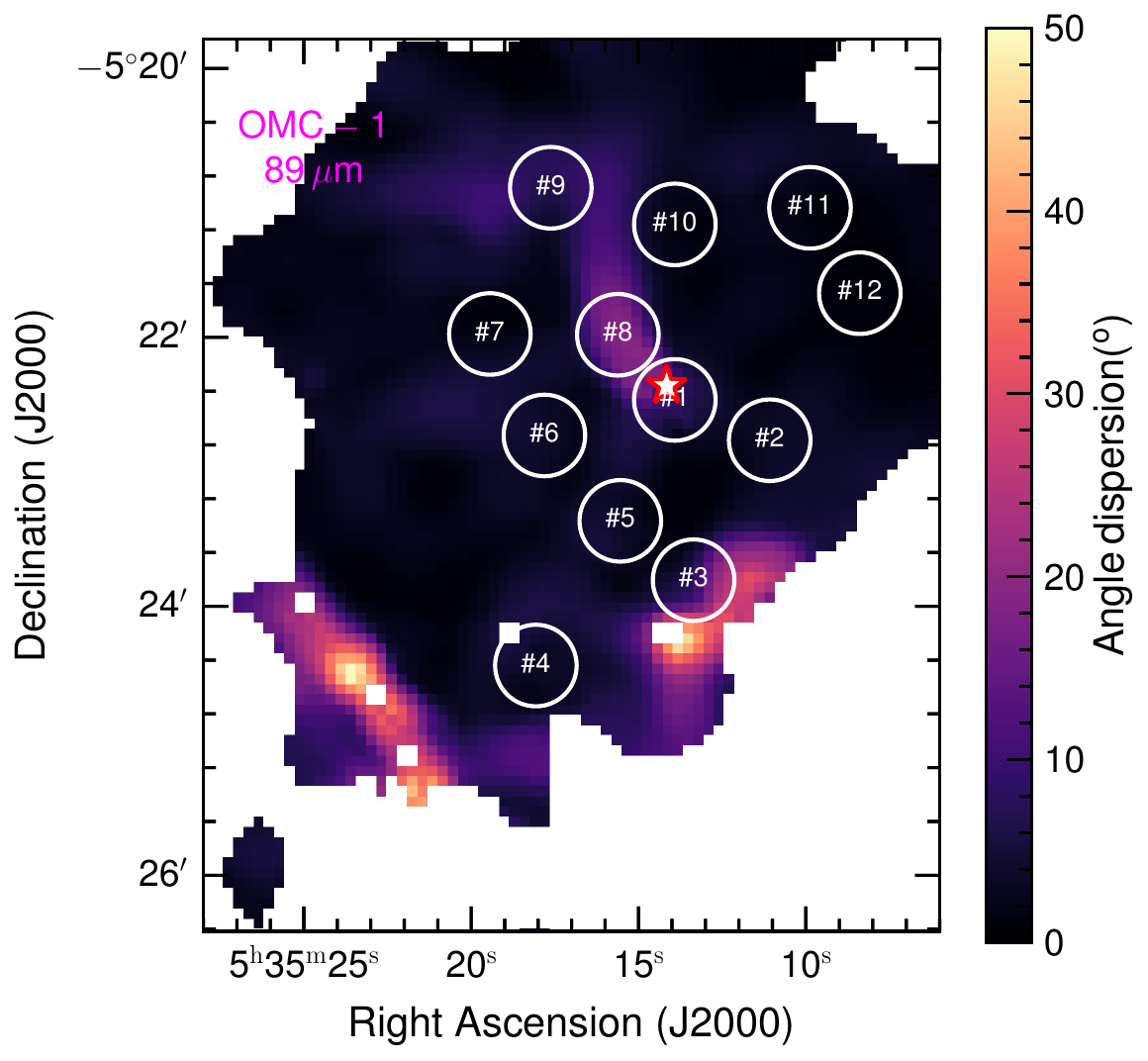}
    \includegraphics[width=0.33\textwidth]{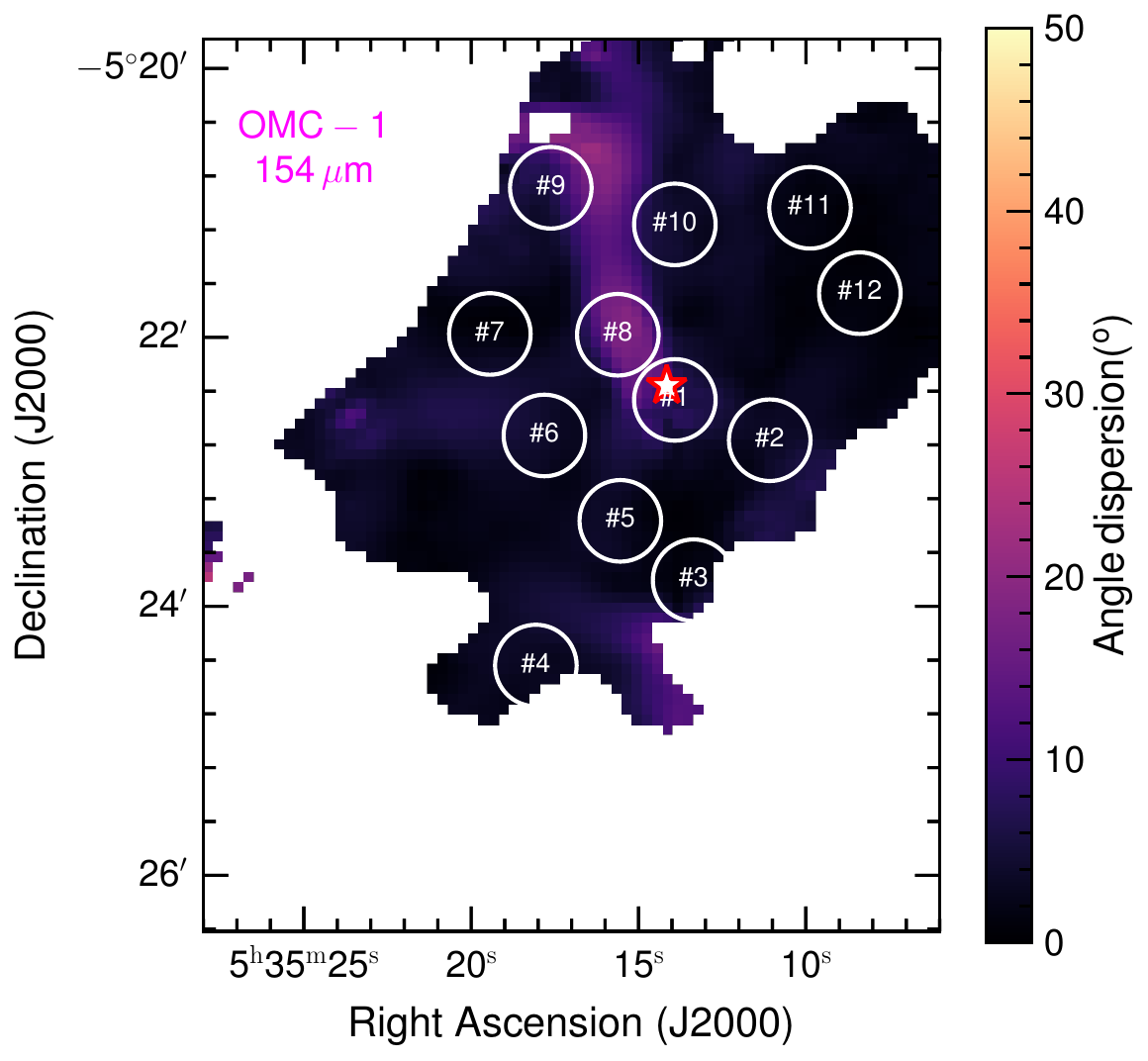}
    \includegraphics[width=0.33\textwidth]{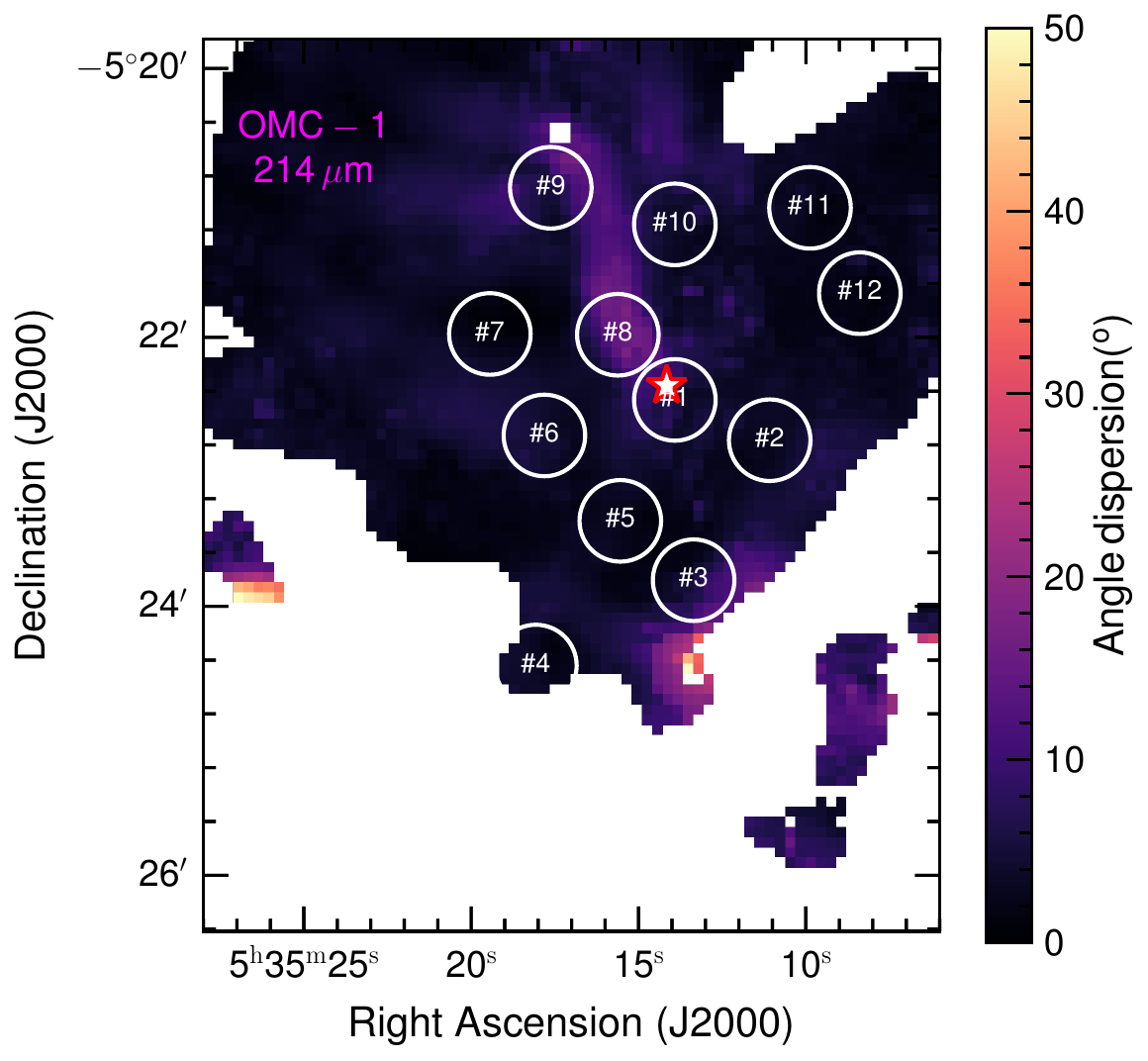}
    \includegraphics[width=0.33\textwidth]{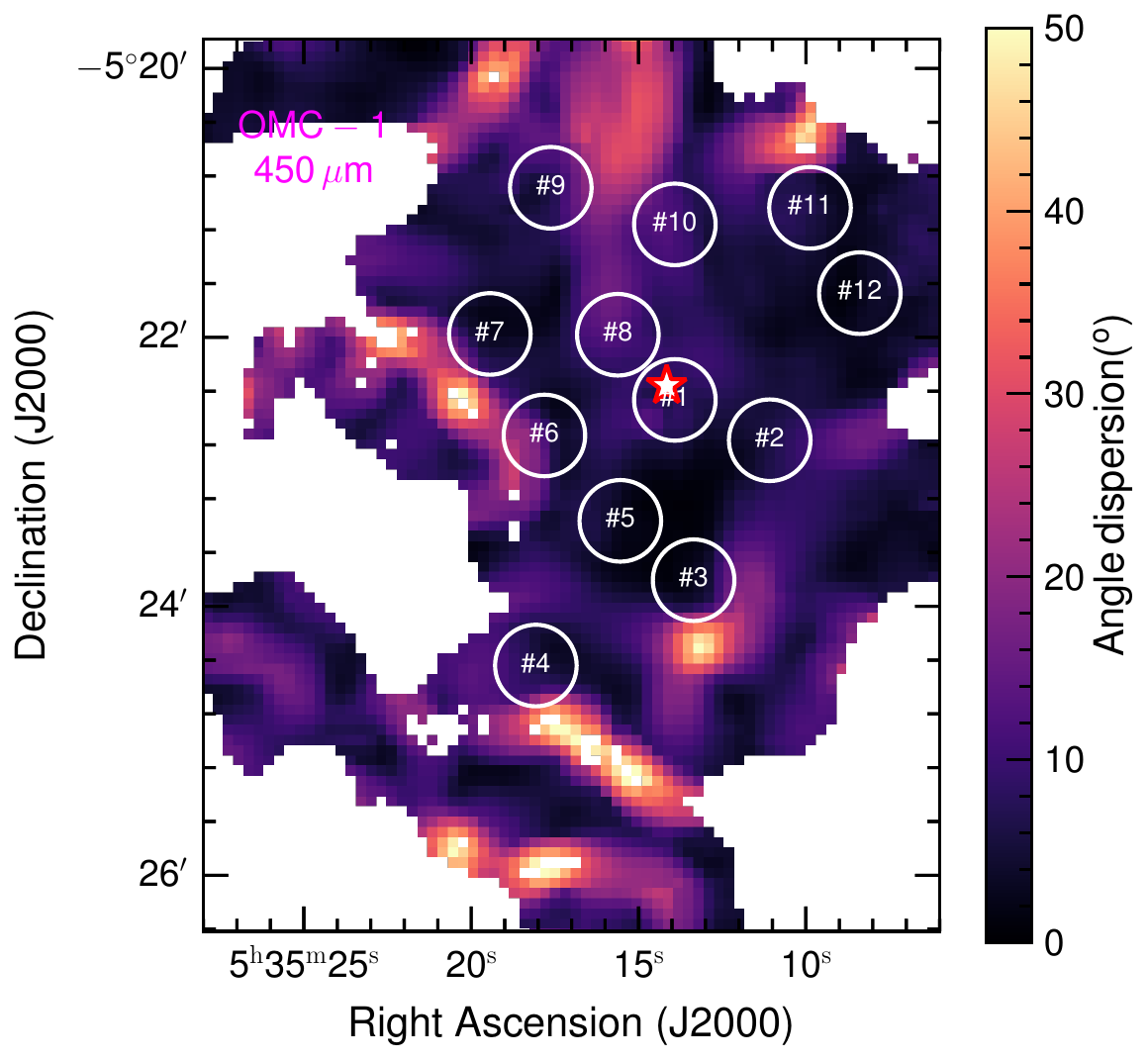}
    \includegraphics[width=0.33\textwidth]{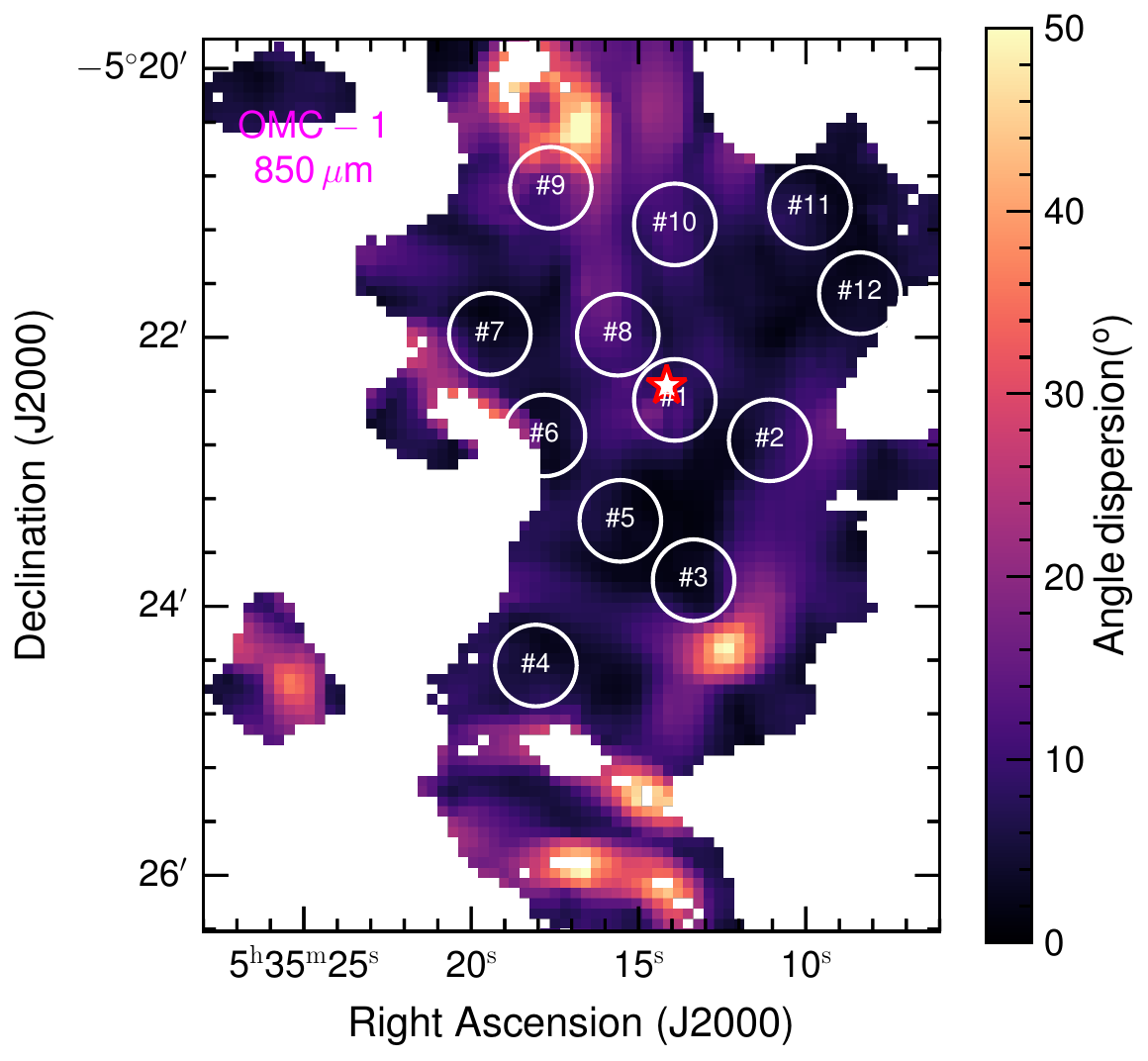}
    \caption{The angle dispersion is observed across all wavelengths. Analysis of submillimetre data indicates that the turbulence areas differ from those seen in the far-infrared data.}
    \label{fig:S}
    \includegraphics[width=0.33\textwidth]{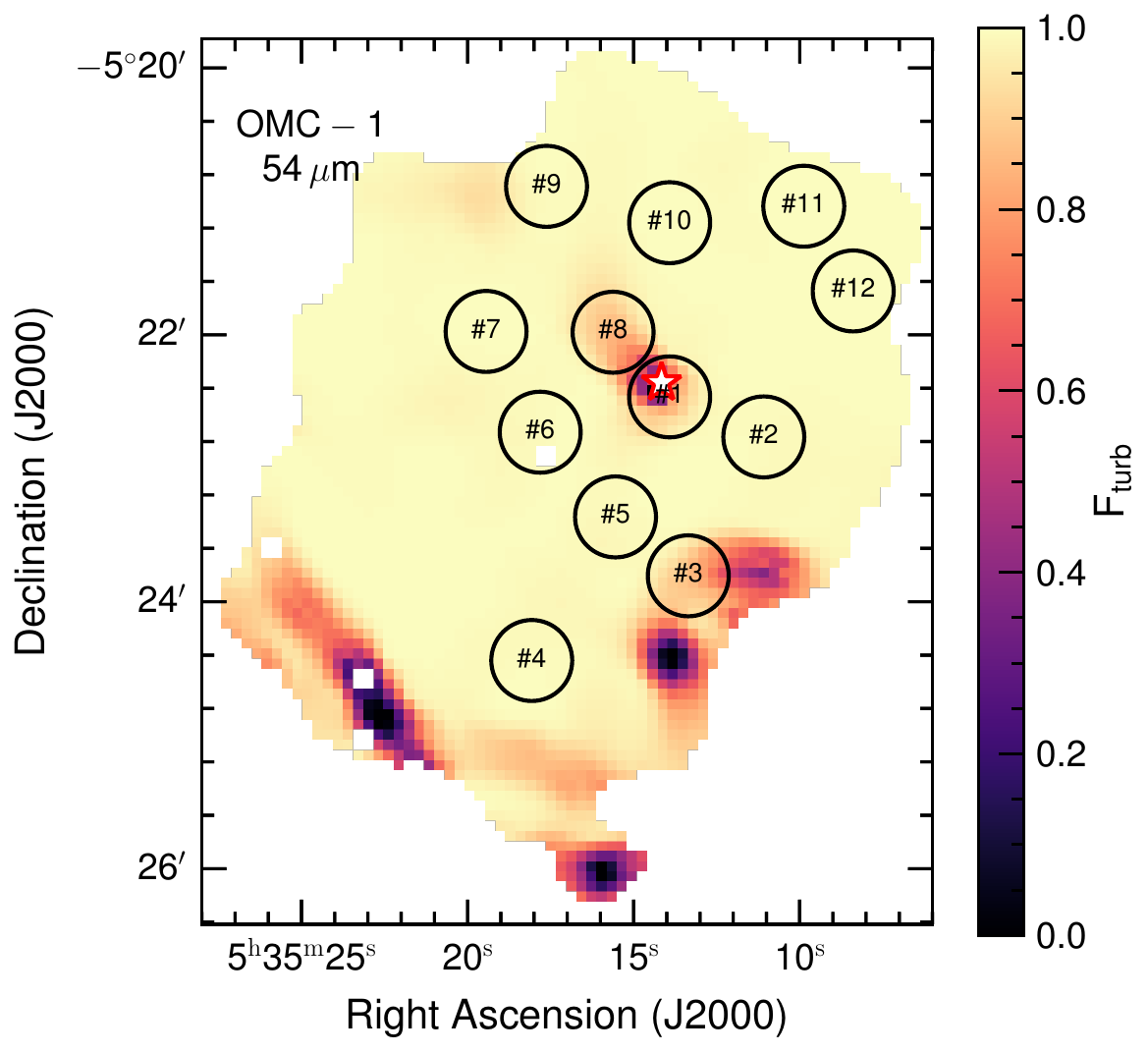}
    \includegraphics[width=0.33\textwidth]{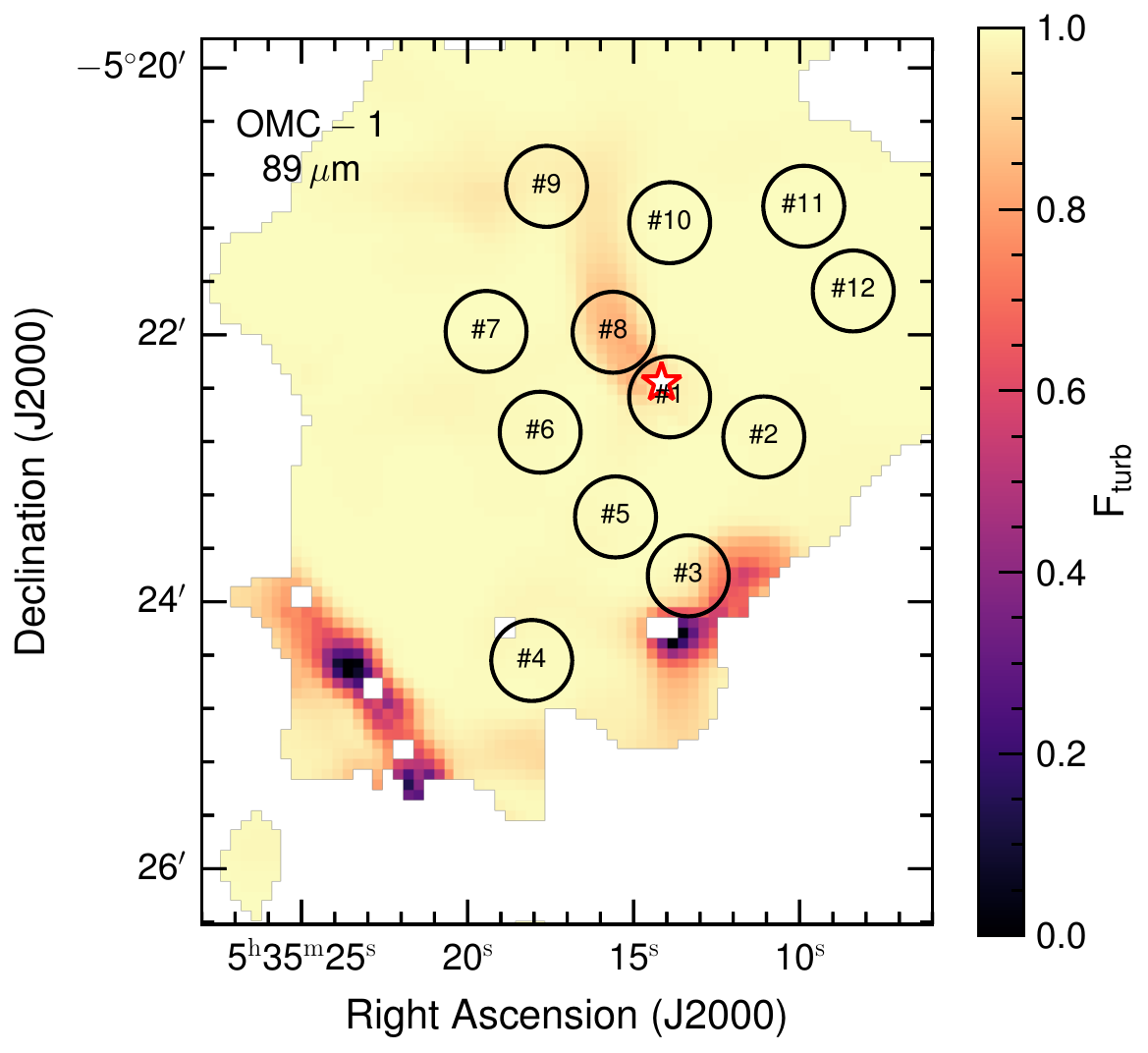}
    \includegraphics[width=0.33\textwidth]{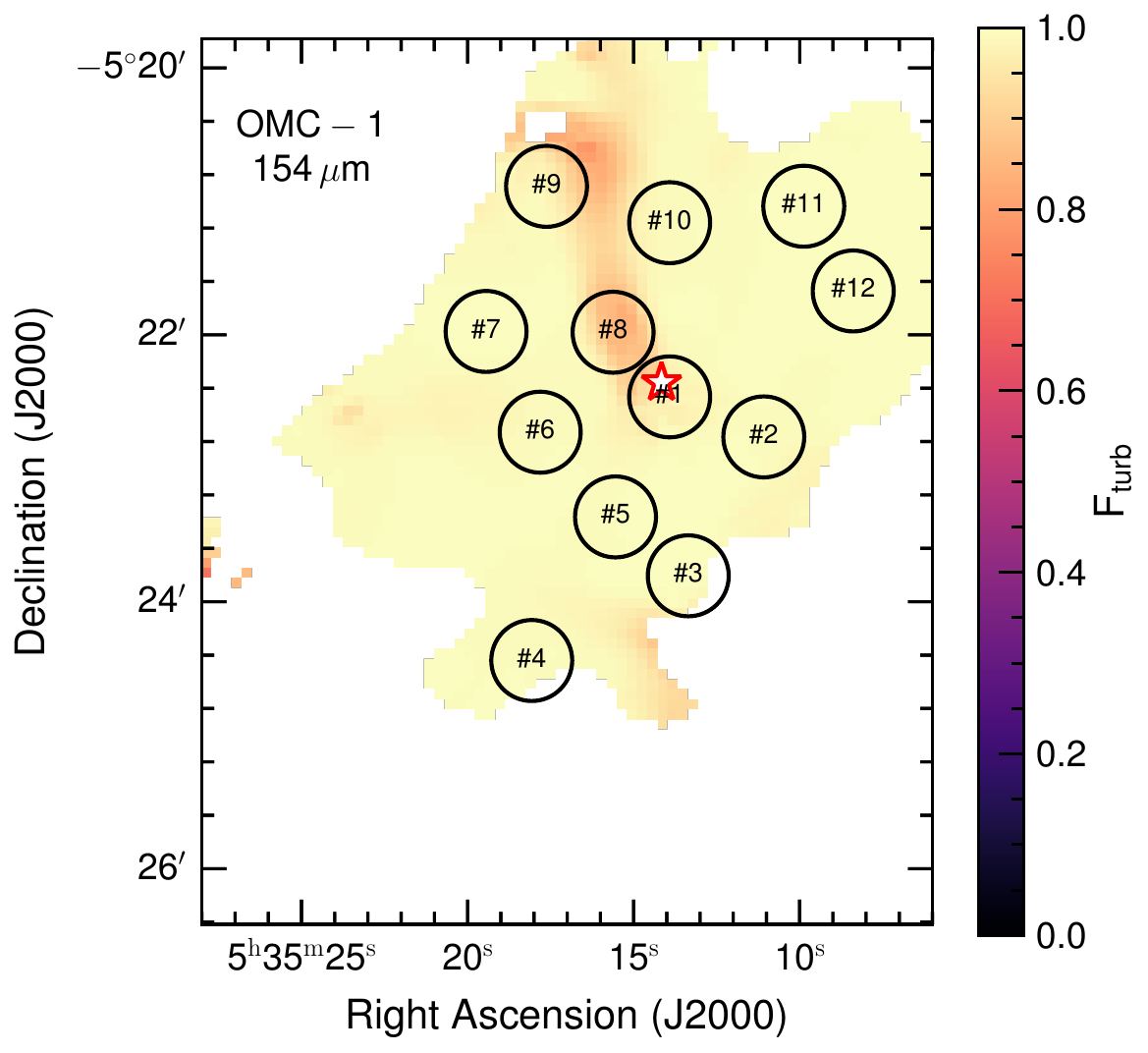}
    \includegraphics[width=0.33\textwidth]{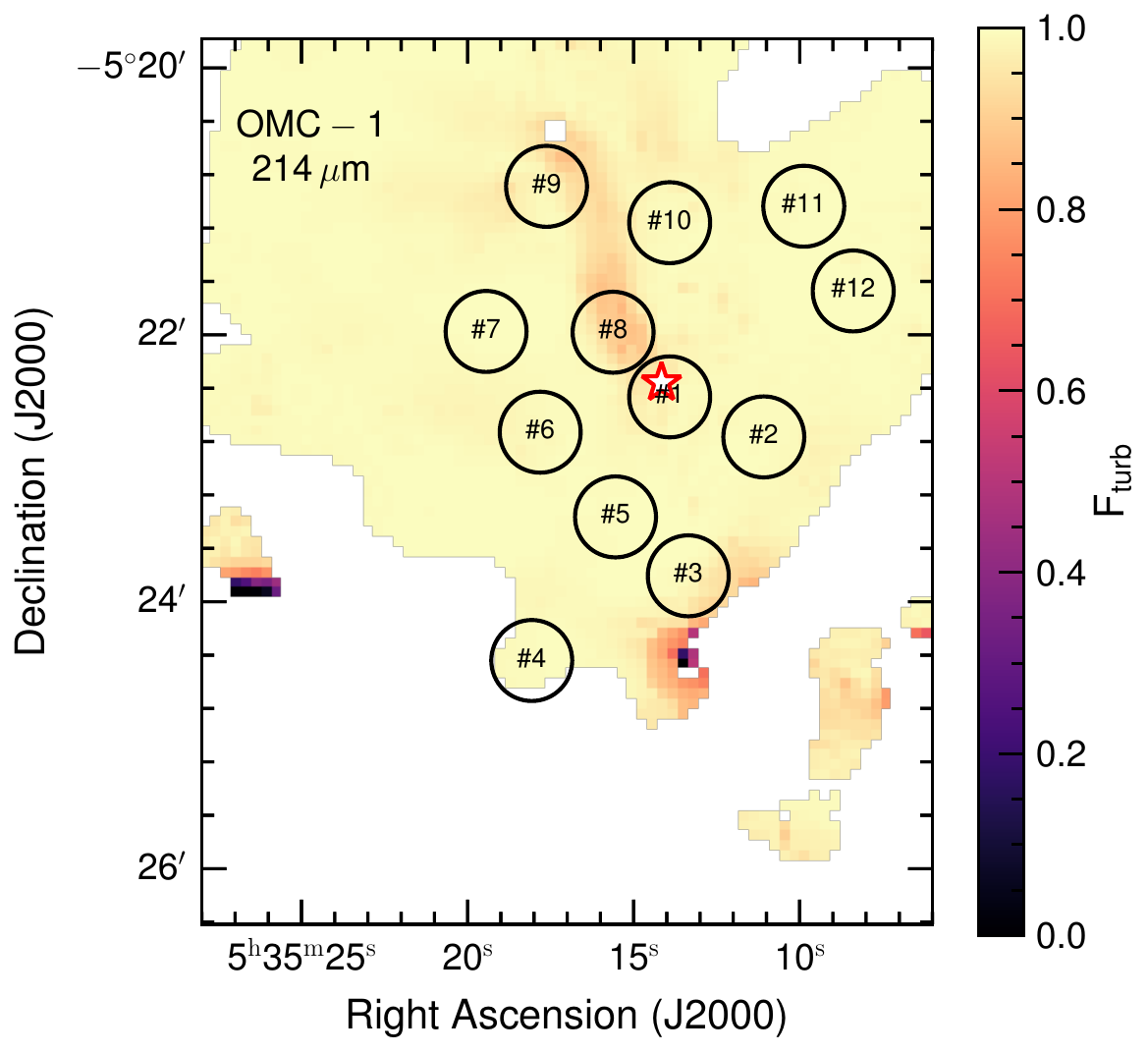}
    \includegraphics[width=0.33\textwidth]{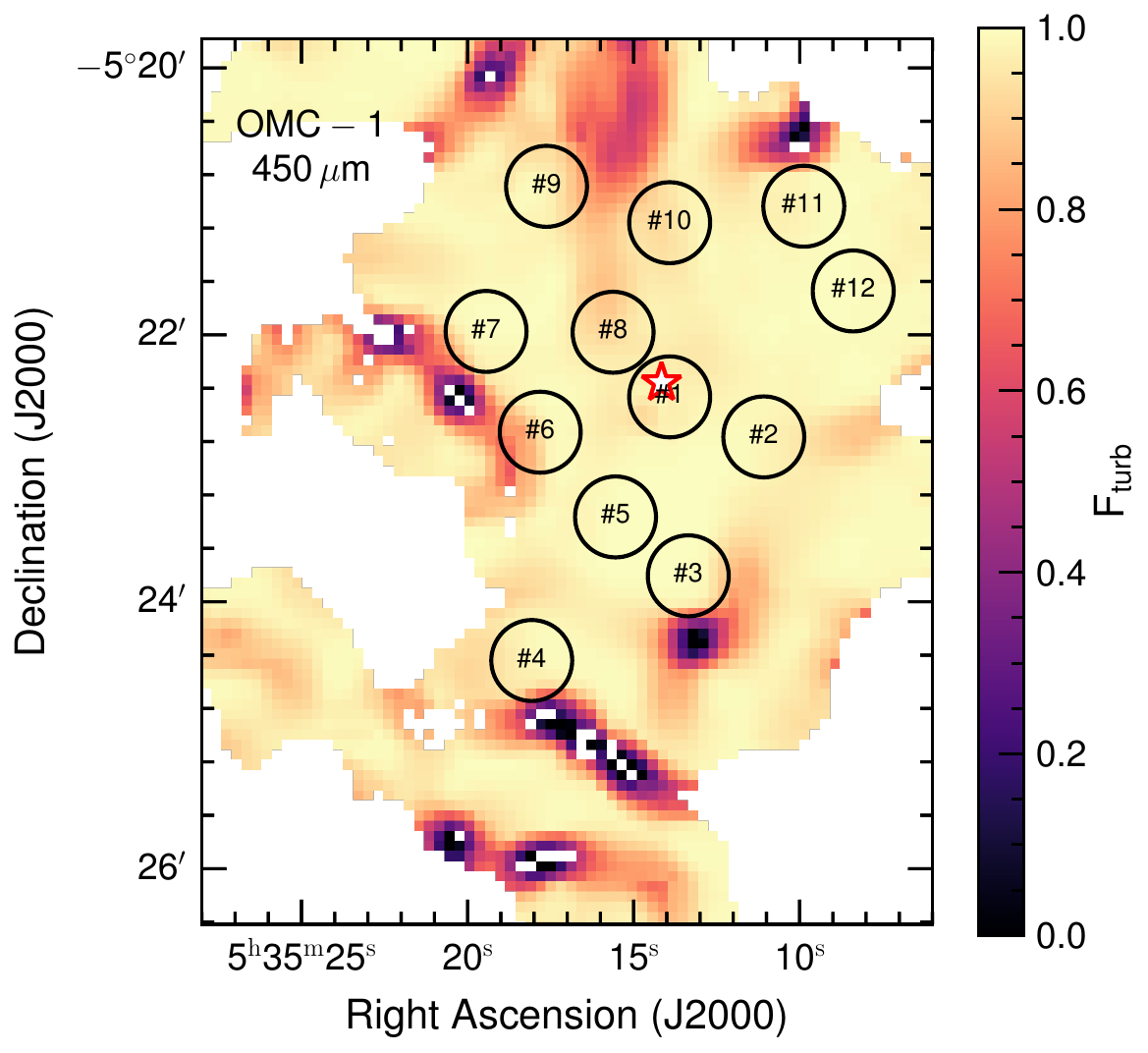}
    \includegraphics[width=0.33\textwidth]{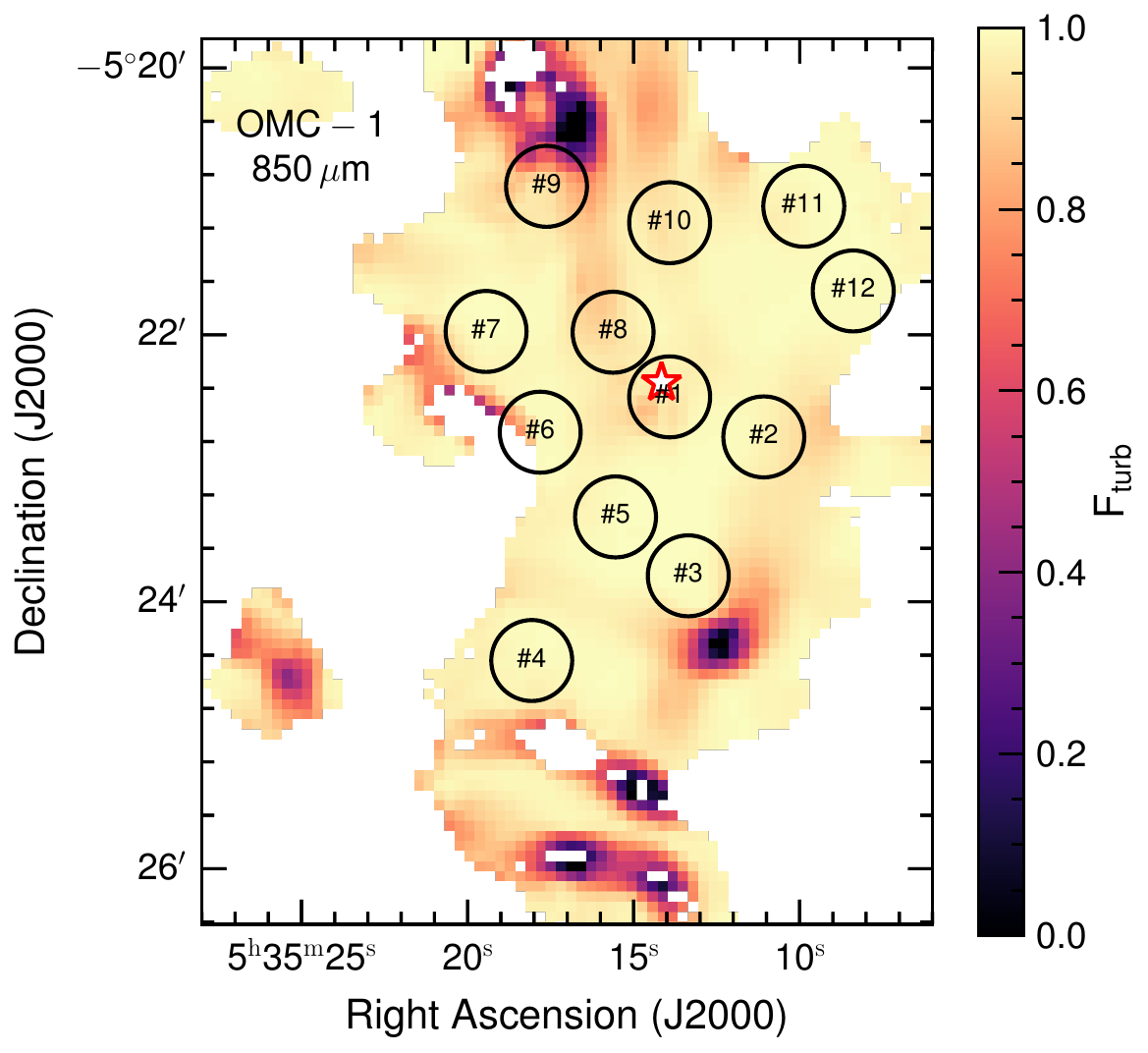}
    \caption{Similar to Figure \ref{fig:S} but for the turbulent parameter $F_{\rm turb}$. The condition Fturb = 1 indicates the absence of turbulence in the magnetic field, while greater turbulence results in a decreased value of $F_{\rm turb}$.}
    \label{fig:Fturb}
\end{figure*}

We qualify the variation of the magnetic field variation on the plane of the sky within the main beam by the dispersion function $S$ (introduced in \citealt{2020A&A...641A..12P}) and turbulent parameter $F_{\rm turb}$ (introduced in \citealt{2023arXiv231017048H}). We refer to Appendix \ref{app:obs_Bfield_tangling} for details of the formulations. The maps of $S$ and $F_{\rm turb}$ for all wavelengths are shown in Figures \ref{fig:S} and \ref{fig:Fturb}.

With a convolved FWHM of $18\rlap{.}''2$, it is clear that the turbulence in the magnetic field is insignificant in most chosen regions, where it is defined by $F_{\rm turb}=1$. However, the variation is strongest at the location of the bar (not included in this study due to lack of available data), the BN/KL core (region \#1) and around the region \#3. It is noteworthy that there are new areas with significant fluctuations in magnetic fields at submillimetre wavelengths, which differ from those observed at far-infrared wavelengths and may support the idea of other layers of dust along the line of sight.   

\subsection{Footprint of Foreground/Background Components}
At certain locations (regions \#2, \#3, \#6, \#7, \#9, \#10, \#11, and \#12 in Figure \ref{fig:modelvsobs_2phase}), the observed polarisation spectrum displayed a V-shaped pattern. Our model, which incorporates two-phase dust (warm and cold phases) at these locations, was able to accurately reproduce these observations. The V-shaped polarisation spectrum could potentially indicate the presence of a cold foreground or background cloud along the observed sight line, as the second phase in our model can exist either in front of or behind the first phase. We propose that utilising multi-wavelength thermal dust polarisation observations could serve as a valuable tool for detecting multi-component dust. This approach offers greater sensitivity compared to the total intensity SED. The primary underlying cause stems from the fact that, while the overall intensity in the cold phase can be outshone by the SED of the warm phase (i.e., see a smooth SED of OMC-1 shown in Figure 2 of \citealt{2019ApJ...872..187C}), it is feasible to differentiate the cold layer, as long as grain alignment is effective, leading to a sufficiently high level of induced polarisation.

It is important to mention that the fits obtained from the two-phase model \textit{do not} exhibit a 'V-shape' in the polarisation spectra in the regions \#1, \#4, \#5, and \#8, which is similar to those obtained from the one-phase models. This indicates that the shape of the polarisation spectrum varies significantly in OMC-1 and is sensitive to local conditions.

\subsection{Potential Evidence for Mixing Compositions of Dust Grains}
Based on our most appropriate model, it was observed that polarisation generation is caused by a mixture of silicate and carbonaceous grains in the cold phase (second phase). As grains could grow efficiently through the accretion of metals to grains at a typical density of $\sim 10^{3}\,\rm cm^{-3}$ (\citealt{2000PASJ...52..585H}) and metal depletion is greater in cold medium (\citealt{1996ARA&A..34..279S}) or coagulation at a higher typical density of $\gtrsim 10^{3}\,\rm cm^{-3}$ (\citealt{2003A&A...398..551S}), the grain composition mixture could potentially be the result of grain growth under cold conditions.

It is crucial to mention that due to the degeneracy of the free parameter $f_{\rm scale}$, the model \textit{cannot} distinguish between different grain compositions (whether silicate alone or carbonaceous in combination) in the first (warm) phase. However, we observed that the model with only silicate composition in the first phase completely fails to reproduce the observed spectra in the regions \#2, \#6 and \#7.

\subsection{Constraining the 3D Magnetic Field Structure from Dust Polarisation}
In our numerical modelling above, we combined the maximum alignment degree and the inclination angle of B-fields into one term, $f_{\rm max}\times \sin^{2}\psi$, and obtained the best-fit parameter for this product.  However, we neglected the impact of variations in the magnetic field orientation along the sight line and within the main beam area (magnetic field tangling).

Here, we discuss the possibility of probing the angle of inclination of the B-field $\psi$ based on the best-fit value of the product $f_{\rm max}\times \sin^{2}\psi$. According to magnetically enhanced RAT (MRAT) alignment theory, grains with embedded iron inclusions can achieve a maximum alignment of $f_{\rm max}=1$ \citep{2016ApJ...831..159H}. The existence of such superparamagnetic grains has been evident from numerous comparisons between numerical models and observations, either using {\it Planck} (\citealt{2023ApJ...948...55H}) or ALMA (\citealt{2023MNRAS.520.3788G}). Therefore, if we assume the most likely model with $f_{\rm max}=1$, we can infer the inclination angle of the B-field based on the best fit values. The results for the $\psi$ angles are shown in Table \ref{tab:Bfield_angle}. 

In general, the value of the magnetic field inclination $\psi$ varies in OMC-1 locally. However, this variation is insignificant in the regions west of the main filament (\#2, \#10, \#11, \#12), and along the main filament (\#1, \#3, \#8) except \#9 where the gas density is lower than in others. Furthermore, it is interesting to note that this angle seems to decrease from the HII region (east of the main filament) to the main filament and further west of the main filament. Therefore, \textit{it suggests that we have a bow-shaped structure of the magnetic field that curves around the filament in OMC-1}. What we have implied is similar to \cite{2019A&A...632A..68T} where the authors showed a bow morphology of the magnetic fields that wrap the OMC-A filament, but on a much larger physical scale. The complete mapping of the angle $\psi$ could potentially reveal the three-dimensional structure of the magnetic field in OMC-1. We note that this approach has already been tested using synthetic polarisation of MHD simulations in \cite{2023arXiv231017048H}, but it is the first time applied to observational data in this paper. 
\begin{table}[!ht]
    \centering
    \caption{The inclination of the magnetic field with respect to the line of sight in the case of $f_{\rm max}=1$ for the two-phase model.} 
    \label{tab:Bfield_angle}     
    \renewcommand{\footnoterule}{} 
    \begin{tabular}{c c c c} 
    \multicolumn{4}{c}{Determination of 3D magnetic field from thermal dust polarisation} \\
    \hline\hline        
    Regions & $f_{\rm max}\times \sin^{2}\psi$ & $\psi(^{o})$ & location\\
    \hline
    \#1     & 0.53 & 47 & Main filament\\
    \#3     & 0.57 & 49 & Main filament\\
    \#8     & 0.58 & 50  & Main filament\\
    \#9     & 0.16 & 24 & Main filament\\
    \#5     & 0.69 & 56 & HII-filament border\\
    \#4     & 0.58 & 50 & HII (East to filament)\\
    \#6     & 0.94 & 76  & HII (East to filament)\\
    \#7     & 0.94 & 76  & HII (East to filament)\\
    \#2     & 0.29 & 33 & West to filament\\
    \#10    & 0.32 & 34 & West to filament\\
    \#11    & 0.34 & 36 & West to filament\\
    \#12    & 0.38 & 38 & West to filament\\
    \hline
    \hline   
    \end{tabular}
\end{table}

\subsection{Caveats} \label{sec:caveats}
The polarisation spectrum towards OMC-1 was constructed using observations from SOFIA/HAWC+ and JCMT/Pol-2. However, our work is subject to various uncertainties, both in terms of observations and computations. The upward slope of the V-shape in Figure \ref{fig:modelvsobs_1phase} is mainly influenced by two observable data points at 450 and 850$\,\mu$m. In certain regions (e.g. regions \#2, \#3, and \#12), the dust polarisation at 850$\,\mu$m is relatively high, resulting in a steep slope towards longer wavelengths. This discrepancy reduces the goodness of fit in these specific regions. To verify this, we compared the 870$\,\mu$m observations and found a discrepancy with the 850$\,\mu$m observations. Therefore, it is crucial to conduct further observations at submillimetre wavelengths (e.g., JCMT/POL-2 at 450 and 850$\,\mu$m, and APEX/A-MKID at 350 and 870$\,\mu$m) and millimetre wavelengths (e.g. IRAM/NIKA-2 at 1.2$\,$mm) to improve the precision of the dust polarisation spectrum. Furthermore, data from SOFIA/HAWC+ and JCMT/POL-2 were processed using different pipeline reductions. These variations in the reduction process could introduce biases in the resulting polarisation degree. It is necessary to develop a consistent procedure for reducing data obtained from different telescopes. As demonstrated in this study, SOFIA/HAWC+ and JCMT/POL-2 observe dust polarisation from different layers along the line of sight. However, it is also possible that there is an additional large-scale magnetic field component present in the SOFIA/HAWC+ observations, which causes the differences between these two observations. A further uncertainty results from the instrumental sensitivity of the polarisation degree to the removal of sky backgrounds.

Our simplified model contains certain uncertainties. First, the model relies on locally observed physical properties of the environment, such as gas density and dust temperature. Therefore, these values are subject to change when a higher spatial resolution is considered. Second, our investigation focusses only on the parameter space of the most important parameters ($\beta$, $f_{\rm max}$, $f_{\rm heat}$, and $f_{\rm scale}$), while other parameters remain unexplored. For example, the dust-to-gas ratio (which we fixed at 1/100), the silicate/carbon ratio (which we fixed at 1.12 as in \citealt{1984ApJ...285...89D}), and the composite grain tensile strength ($S_{\rm max}=10^{7}\,\rm erg\,cm^{-3}$). Different constrained best models may arise from variations in these parameters.

In addition, our assumption is that the emissions in the FIR/Submm range are optically thin in both the cold and warm phases along the line of sight. Therefore, our model \textit{cannot} accurately describe radiative processes and their impact throughout these phases. To address this issue, a more accurate modelling approach can be utilised that incorporates a proper radiative transfer process and the RAT paradigm, such as POLARIS+ in \cite{2023MNRAS.520.3788G}.

\section{Conclusions}\label{sec:Conclusion}
In this study, we presented the polarisation of thermal dust emission in OMC-1 using data from four bands of SOFIA/HAWC+ (54, 89, 154, and 214$\,\mu$m) and two bands of JCMT/POL-2 (450 and 850$\,\mu$m). In general, our analysis revealed that the slope of the polarisation spectrum is positive along the main filament structure and the Orion Bar, where the gas density is relatively high. In regions with lower gas density, the slope is negative. 

We compared our simplified model based on the RAT paradigm and accounted for the inclined magnetic field with respect to the line of sight with the dust polarisation spectra observed in OMC-1. Although the best-fit models were able to reproduce the observations reasonably well in some locations, they failed to explain the pronounced V-shape (the polarisation degree first decreases and then increases towards a longer wavelength) observed in certain spectra. This V-shape suggests the presence of multiple dust layers along the line of sight, which is consistent with the findings of \cite{2022ApJ...930L..22R}. To improve our model, we introduced a 'two-phase' approach that accounts for two dust layers (warm and cold), and this modification resulted in a better fit to the observations across OMC-1. Our analysis also indicated that the dust polarisation spectrum is more sensitive to the presence of multiple dust layers than the total intensity spectral energy distribution. Furthermore, the best-fit model suggested a composite composition of dust grains, consisting of a mixture of silicate and carbonaceous materials, in the cold dust layer.

We showed that the variations of the polarisation angle on the plane of sky within the beam are distinct from submillimetre to FIR wavelengths. This discrepancy likely infers that FIR and submillimetre polarimetry probe different layers of dust along the light of sight. We discussed that if validated, combining the RAT-paradigm theories with the degree of dust polarisation observations may allow inferring the local three-dimensional structure of the magnetic field along the individual line of sight.

We acknowledge that our study has uncertainties that arise from both the observations and the simplicity of our modelling approach. Nevertheless, our work emphasises the importance of utilising multi-wavelength dust polarisation to investigate the mechanisms of grain alignment and dust polarisation, along with the effectiveness of the RAT paradigm in interpreting such polarisation spectra. Our simplified model, a \textsc{python}-based, is publicly available at \href{https://github.com/lengoctram/DustPOL-py}{https://github.com/lengoctram/DustPOL-py}. 

\begin{acknowledgement}
We express our gratitude to the anonymous referee for their detailed comments and suggestions, which have enhanced the scientific impact of this work. We thank Dr. Jihye Hwang for sharing and discussing the thermal dust polarisation from JCMT/POL-2. We also thank Dr. Joseph Michail and Prof. David Chuss for sharing the thermal dust polarisation from SOFIA/HAWC+ and the dust temperature and gas column density maps derived from the dust SED fitting. This research is based on observations made with the NASA/DLR Stratospheric Observatory for Infrared Astronomy (SOFIA) and the James Clerk Maxwell Telescope (JCMT). SOFIA is jointly operated by the Universities Space Research Association, Inc. (USRA), under NASA contract NNA17BF53C, and the Deutsches SOFIA Institut (DSI) under DLR contract 50 OK 0901 to the University of Stuttgart. JCMT is operated by the East Asian Observatory on behalf of The National Astronomical Observatory of Japan, Academia Sinica Institute of Astronomy and Astrophysics in Taiwan, the Korea Astronomy and Space Science Institute, the National Astronomical Observatories of China, and the Chinese Academy of Sciences (grant No.
XDB09000000), with additional funding support from the Science and Technology Facilities Council of the United Kingdom and participating universities in the United Kingdom and Canada.
\end{acknowledgement}

\bibliographystyle{aa} 
\bibliography{bib}

\appendix
\section{Observations}\label{app:obs}
\subsection{Synthesising data} \label{app:obs_syn}
In Section \ref{sec:obs}, we introduce that we worked on the polarisation data at six distinct wavelengths. Thus, we synthesised all Stokes I, Q and U and their associated errors ($\sigma_{I}$, $\sigma_{Q}$, and $\sigma_{U}$) to a common FWHM of 18.2'' and a pixel size of 4.55''. For the APEX/PolKA data, we applied the selection criteria of $I>0$, $I/\sigma_I\geq 10$, and $p/\sigma_p\geq 3$ and dentified only six regions that are covered by both PolKA and HAWC+ and POL-2. Subsequently, we computed the degree of polarisation ($p(\%)$) as follows (see \citealt{2018arXiv181103100G} for details).

The polarized intensity is defined as
\begin{equation}
    I_{p} = \sqrt{Q^{2}+U^{2}} ~~~.
\end{equation} 
Its associated error derived from error propagation, assuming that the errors in Q and U are uncorrelated, is
\begin{equation}
    \sigma_{I_{p}} = \left[\frac{(Q\sigma_{Q})^{2}+(U\sigma_{U})^{2}}{Q^{2}+U^{2}} \right]^{1/2} ~~~.
\end{equation}

The debiased polarisation degree and its associated error are calculated as
\begin{equation}
\begin{split}
    & p_{\rm biased} = \frac{100}{I}\sqrt{Q^{2}+U^{2}} ~~~(\%), \\ 
    & \sigma_{p} = p\left[\left(\frac{\sigma_{I_{p}}}{I_{p}}\right)^{2} + \left(\frac{\sigma_{I}}{I}\right)^{2}\right]^{1/2} ~~~(\%), \\
    & p = \sqrt{p^{2}_{\rm biased} - \sigma^{2}_{p}} ~~~(\%),
\end{split}
\end{equation} 

The polarisation angle and its associated error are defined as
\begin{equation}
\begin{split}
    & \theta = \frac{90}{\pi}{\rm atan2}(U,Q) ~~~(^{o}), \\
    & \sigma_{\theta} = \frac{90}{\pi (Q^{2}+U^{2})}\sqrt{(Q\sigma_{U})^{2} + (U\sigma_{Q})^{2}} ~~~(^{o}).
\end{split}
\end{equation}

\subsection{Characterization of magnetic field tangling}\label{app:obs_Bfield_tangling}
The variation of B-fields could induce depolarisation of thermal dust emission. Thus, we seek the relationship of the degree of polarisation with the dispersion function $S$ of the polarisation angle. The biased dispersion at position \vec{r} on the sight line is given as
\begin{equation}
    S(\vec{r}) = \sqrt{\frac{1}{N}\sum^{N}_{i=1}[\theta(\Vec{r}+\vec{\delta}_{i})-\theta(\vec{r})]^{2}}
\end{equation}
and the associated error is (see Equation 8 in \citealt{2020A&A...641A..12P})
\begin{equation}
    \begin{split}
        \sigma_{S}^{2}(\vec{r}) = & \frac{1}{N^{2}S^{2}}\sum_{i=1}^{N}\sigma^{2}_{\psi}(\vec{r}+\vec{\delta_{i}})[\theta(\vec{r}+\vec{\delta}_{i})-\theta(\vec{r})]^{2} \\
        & +\frac{\sigma^{2}_{\theta}(\vec{r})}{N^{2}S^{2}}\left[\sum_{i=1}^{N}\theta(\vec{r}+\vec\delta_{i}) - \theta(\vec{r})\right]^{2}
    \end{split}
\end{equation}
where $\theta(\Vec{r})$ and $\sigma(\vec{r})$ are the polarisation angle and the associated error at position $\vec{r}$, respectively. Similarly, $\theta(\Vec{r}+\Vec{\delta}$) and $\sigma(\vec{r}+\vec{\delta})$ for position $\vec{r}+\vec{\delta}$. $N$ is the number of angles chosen. In practice, for a given position $\vec{r}$, we select all data points within a circle centred on this position with a diameter of two beam-size. Finally, the debiased dispersion function is computed as $\delta \theta = \sqrt{S^{2}-\sigma^{2}_{S}}$. The distribution of the dispersion angle is shown in Figure \ref{fig:S}.

Another quantity that can qualitatively assess the turbulence in a magnetic field is $F_{\rm turb}$ introduced in \cite{2023arXiv231017048H} as
\begin{equation}
    F_{\rm turb} = \frac{1}{2}\left[3 \langle \cos^{2}(\Delta \theta) \rangle -1 \right]
\end{equation}
with $\Delta \theta$ the deviation in the angle between the local magnetic field and the mean field. Furthermore, \cite{2021ApJ...913...85H} showed that the OMC-1 region is sub-Alfvénic with a mean Alfvénic Mach number of approximately 0.4. Therefore, we can derive the link between $F_{\rm turb}$ and the angle dispersion as (see Eq. 32 in \citealt{2023arXiv231017048H}) 
\begin{equation}
    F_{\rm turb} \simeq 1-1.5(\delta \theta)^{2}
\end{equation}
The condition $F_{\rm turb}=1$ indicates the absence of turbulence in the magnetic field, while greater turbulence results in a decreased value of $F_{\rm turb}$. The distribution of $F_{\rm turb}$ for all wavelengths is shown in Figure \ref{fig:Fturb}.

\section{Modelling}\label{app:model}
\subsection{Recall the one-phase model basics}\label{app:model_onephase}
This study is based on the assumption of uniform and plane-of-sky ambient magnetic fields and does not take into account radiative transfer. Thus, the total and polarized intensity can be estimated analytically, as described by \cite{2020ApJ...896...44L} and \cite{2021ApJ...906..115T}. Our model is based on the theory of the RAT paradigm, and its basic concepts are briefly described below.

The degree of thermal dust polarisation is defined as
\begin{equation}
    p^{\rm ideal}(\%) = 100 \times \frac{I_{\rm pol}}{I_{\rm em}} ~~~{.}    
\end{equation}
Where $I_{\rm pol}$ and $I_{\rm em}$ are the polarized and total intensity of the thermal dust emission. The inclined magnetic field with respect to the line of sight can lower the polarized intensity and then reduce the net degree of the dust polarisation as
\begin{equation}
    p(\%) = p^{\rm ideal}(\%) \times \sin^{2}\psi ~~~{.}
\end{equation}
where $\psi$ is the mean angle of inclination of the regular magnetic field to the line of sight. If there is no component of the magnetic field on the sight line ($\psi=90^{0}$), the predicted $p(\%)$ is the maximum as $p^{\rm ideal}(\%)$.

As we only consider a dusty environment containing carbonaceous and silicate grains, the total emission intensity is
\begin{equation}
    \frac{I^{\rm 1st-phase}_{\rm em}(\lambda)}{N_{\rm H}} = \sum_{j=\rm sil,car} \int^{a_{\rm max}}_{a_{\rm min}} Q_{\rm ext}\pi a^{2} 
    \times \int_{T_{d}} dT B_{\lambda}(T_{\rm d})\frac{dP}{dT}\frac{1}{n_{\rm H}}\frac{dn_{j}}{da}da.~~~
\end{equation}
where $B_{\lambda}(T_{\rm d})$ is the black-body radiation at dust temperature $T_{\rm d}$, $dP/dT$ is the distribution of dust temperature, Q$_{ext}$ is the extinction coefficient, $dn/da \sim a^{-\beta}$ is the grain-size distribution. As we work on the thermal dust polarisation, we adopt the MRN-like distribution with the power index as a free parameter. The dust temperature distribution depends on the grain size and radiation strength, which is computed by the DustEM code (\citealt{2011A&A...525A.103C}, see, e.g. Figure 8 in \citealt{2020ApJ...896...44L}).

If silicate and carbon are separated populations, the silicates can align with the ambient magnetic field, while the carbon grains cannot. The polarized intensity is given by
\begin{equation}
    \frac{I^{\rm 1st-phase}_{\rm pol}(\lambda)}{N_{\rm H}}= \int^{a_{\rm max}}_{a_{\rm align}} f(a)Q_{\rm pol}\pi a^{2}
    \times\int dT B_{\lambda}(T_{\rm d})\frac{dP}{dT}\frac{1}{n_{\rm H}}\frac{dn_{\rm sil}}{da}da, ~~~
\end{equation}
$Q_{\rm pol}$ is the polarisation coefficient and $f(a)$ is the alignment function. Size-dependent $f(a)$ is empirically described as
\begin{equation}
    f(a) = f_{\rm max}\left[1-e^{-(0.5a/a_{\rm align})^{3}} \right]
\end{equation}
with $a_{\rm align}$ the alignment size, above which grain is aligned. $f_{\rm max}$ is the maximum alignment efficiency. $f_{\rm max}=1$ stands for perfect alignment. It indicates that the grains with $a \geq a_{\rm align}$ align with an efficiency of $f_{\rm max}$, while the grains will not align otherwise. We varied the product of $f_{\rm max}\times \sin^{2}\psi$ as a free parameter. The alignment size $a_{\rm align}$ is determined by a condition in which the angular velocity obtained by RATs ($\omega_{\rm RAT}(a)$) is three times the thermal angular velocity ($\omega_{\rm th}$), which yields (see, e.g. \citealt{2021ApJ...908..218H} for the analytical formulations of $\omega_{\rm RAT}(a)$ and $\omega_{\rm th}$)
\begin{equation}
    \begin{split}
        a_{\rm align} \simeq & 0.055 \left(\frac{\rho}{3\,\rm g\,cm^{-3}}\right)^{-1/7}\left(\frac{\gamma}{0.1}\right)^{-2/7}\left(\frac{n_{\rm H}}{10^{3}\,\rm cm^{-3}}\right)^{2/7}U^{-2/7} \\
        & \times \left(\frac{T_{\rm gas}}{10\,\rm K}\right)^{2/7} \left(\frac{\bar{\lambda}}{1.2\,\rm \mu m}\right)^{4/7}(1+F_{\rm IR})^{2/7} ~~~ {\rm \mu m}
    \end{split}
\end{equation}
where $\rho$ is the volume mass density of grain, $\gamma$ is the degree of radiation anisotropy, $\bar{\lambda}$ is the mean wavelength of the radiation field, and $F_{\rm IR}$ is the ratio of the damping rate caused by infrared emission to the one caused by gas collisions. In this work, we fix $\gamma=1$, which stands for the unidirectional radiation field. $\gamma=0.1$ for diffuse medium and 0.7 for molecular cloud (\citealt{1996ApJ...470..551D}), while it is 0.3 in dense core (\citealt{2007ApJ...663.1055B}). In proximity to the luminous source (protostars), \cite{2023MNRAS.520.3788G} showed that $\gamma \simeq 1$.

If silicate and carbon grains are mixed, which may exist in dense clouds due to many cycles of photoprocessing, coagulation, shattering, accretion, and erosion (see e.g. \citealt{2013A&A...558A..62J}), the carbon grains could passively align with the ambient magnetic field due to this mixture and their thermal emission could be polarised. For the simplest case, assuming these grain populations have the same alignment parameters (i.e., $a_{\rm align}$, $f(a)$), the total polarized intensity is
\begin{equation}
    \begin{split}
        \frac{I^{\rm 1st-phase}_{\rm pol}(\lambda)}{N_{\rm H}} = & \sum_{j=\rm sil,car} \int^{a_{\rm max}}_{a_{\rm align}} f(a)Q_{\rm pol}\pi a^{2} \\
        & \times\int_{T_{\rm d}} dT B_{\lambda}(T_{\rm d})\frac{dP}{dT}\frac{1}{n_{\rm H}}\frac{dn_{j}}{da}da.
    \end{split}
\end{equation}
The extinction and polarisation coefficients are calculated by the DDSCAT model (\citealt{1994JOSAA..11.1491D, 2008JOSAA..25.2693D}; \citealt{2012OExpr..20.1247F}) for a prolate spheroidal grain shape with an axial ratio of 1/3.

\subsection{Two-phase model}\label{app:model_twophase}
We follow that approach in \cite{2023arXiv231017211S}, the relative difference in intensities between the two phases is characterized by a factor $f_{\rm scale}$, the relative abundance between two phases is characterized by a factor $f_{\rm scale,car}$. The intensity of the radiation (equivalent to the temperature of the dust) in the second phase is related to that of the first phase as $f_{\rm heat}=T^{\rm 1st-phase}_{\rm d}/T^{\rm 2nd-phase}_{\rm d}$. In this simplified model, we made two critical assumptions. First, the power index of the distribution in the second phase is the same as in the first phase. Second, the total and polarized intensity are optically thin. With these assumptions, the total thermal dust intensity is a summation over two phases, as
\begin{equation}
\begin{split}
    I^{\rm total}_{\rm em} = & I^{\rm 1st-phase}_{\rm em} \\ 
    & + f_{\rm scale}\times \int^{a_{\rm 2,max}}_{a_{\rm 2,min}} Q_{\rm ext}\pi a^{2} \\
    & \times \int_{T_{\rm d}} dT B_{\lambda}(T^{\rm sil,2}_{\rm d})\frac{dP_{\rm sil,2}}{dT}\frac{1}{n_{\rm H}}\frac{dn_{\rm sil,2}}{da}da \\ 
    & + f_{\rm scale}\times f_{\rm scale,car}\times \int^{a_{\rm 2,max}}_{a_{\rm 2,min}} Q_{\rm ext}\pi a^{2} \\
    & \times \int_{T_{\rm d}} dT B_{\lambda}(T^{\rm car,2}_{\rm d})\left(\frac{dP_{\rm car,2}}{dT}\right)\frac{1}{n_{\rm H}}\frac{dn_{\rm car,2}}{da}da.~~~
    \end{split}
\end{equation}

If only silicate is considered to be aligned in the second phase, the total polarized intensity is
\begin{equation}
    \begin{split}
    I^{\rm total}_{\rm pol} = & I^{\rm 1st-phase}_{\rm pol} \\ 
    & + f_{\rm scale}\times \int^{a_{\rm max}}_{a_{\rm align}} f(a)Q_{\rm pol}\pi a^{2} \\
    & \times\int_{T_{\rm d}} dT B_{\lambda}(T^{\rm sil,2}_{\rm d})\frac{dP_{\rm sil,2}}{dT}\frac{1}{n_{\rm H}}\frac{dn_{\rm sil,2}}{da}da,
    \end{split}
\end{equation}

If a mixture of silicate and carbonaceous grains are aligned, the total polarized intensity is then
\begin{equation}
    \begin{split}
        I^{\rm total}_{\rm pol} = & I^{\rm 1st-phase}_{\rm pol} \\
        & + f_{\rm scale}\times \int^{a_{\rm max}}_{a_{\rm align}} f(a)Q_{\rm pol}\pi a^{2} \\
        & \times\int_{T_{\rm d}} dT B_{\lambda}(T^{\rm sil,2}_{\rm d})\frac{dP_{\rm sil,2}}{dT}\frac{1}{n_{\rm H}}\frac{dn_{\rm sil,2}}{da}da \\
        & + f_{\rm scale}\times f_{\rm scale,car}\times \int^{a_{\rm max}}_{a_{\rm align}} f(a)Q_{\rm pol}\pi a^{2} \\
        & \times\int_{T_{\rm d}} dT B_{\lambda}(T^{\rm car,2}_{\rm d})\frac{dP_{\rm car,2}}{dT}\frac{1}{n_{\rm H}}\frac{dn_{\rm car,2}}{da}da
    \end{split}
\end{equation}
where the subscript and superscript number 2 refer to the second phase. In this work, we \textit{do not} consider the variation in abundance of the carbonaceous grain between two phases and thus keep $f_{\rm scale,car}=1$ as a constant. Another practical reason for not varying $f_{\rm scale,car}$ is the degeneration of the fitting given 6 data points in the spectrum.

\subsection{Modelling the polarisation spectrum and fit to observations}\label{app:model_fitting}
\begin{figure*}[!ht]
    \centering
    \includegraphics[width=0.45\textwidth]{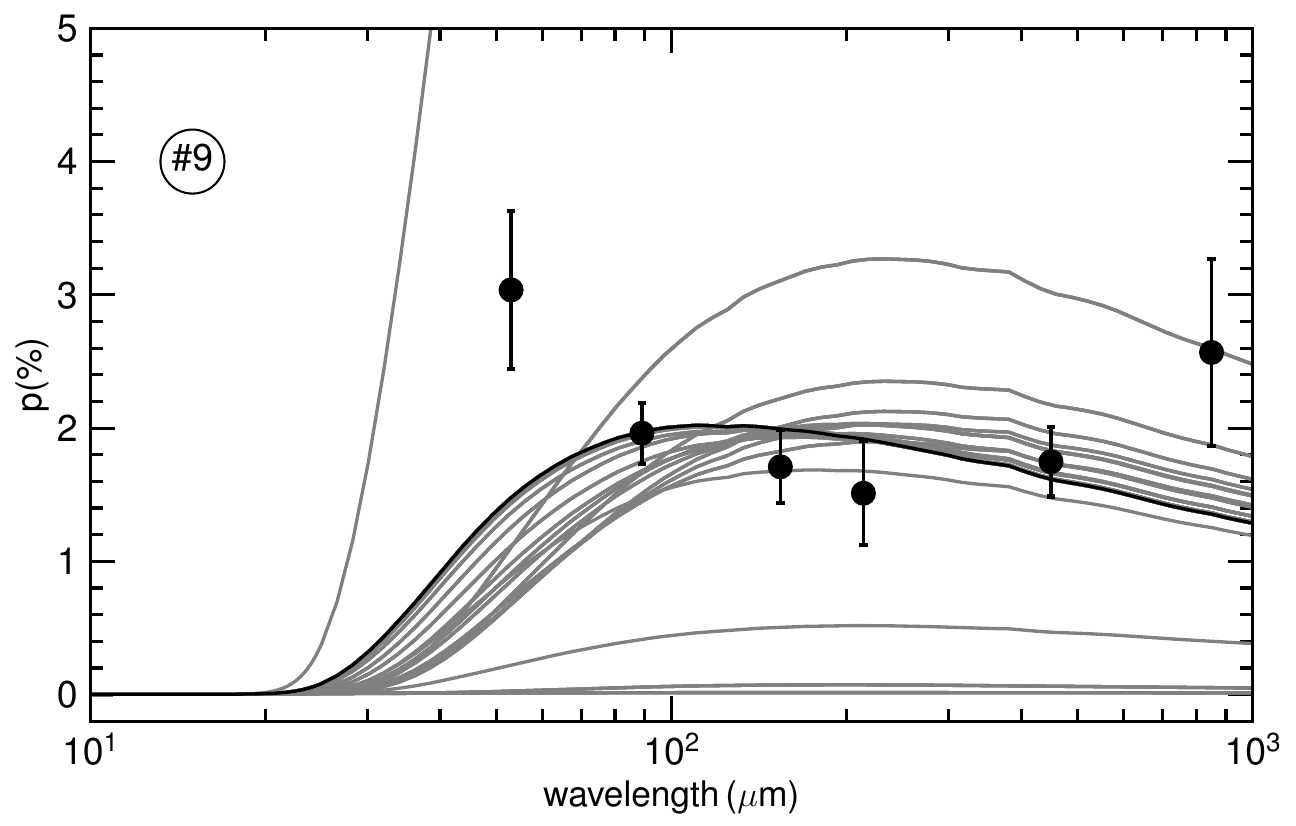}
    \includegraphics[width=0.45\textwidth]{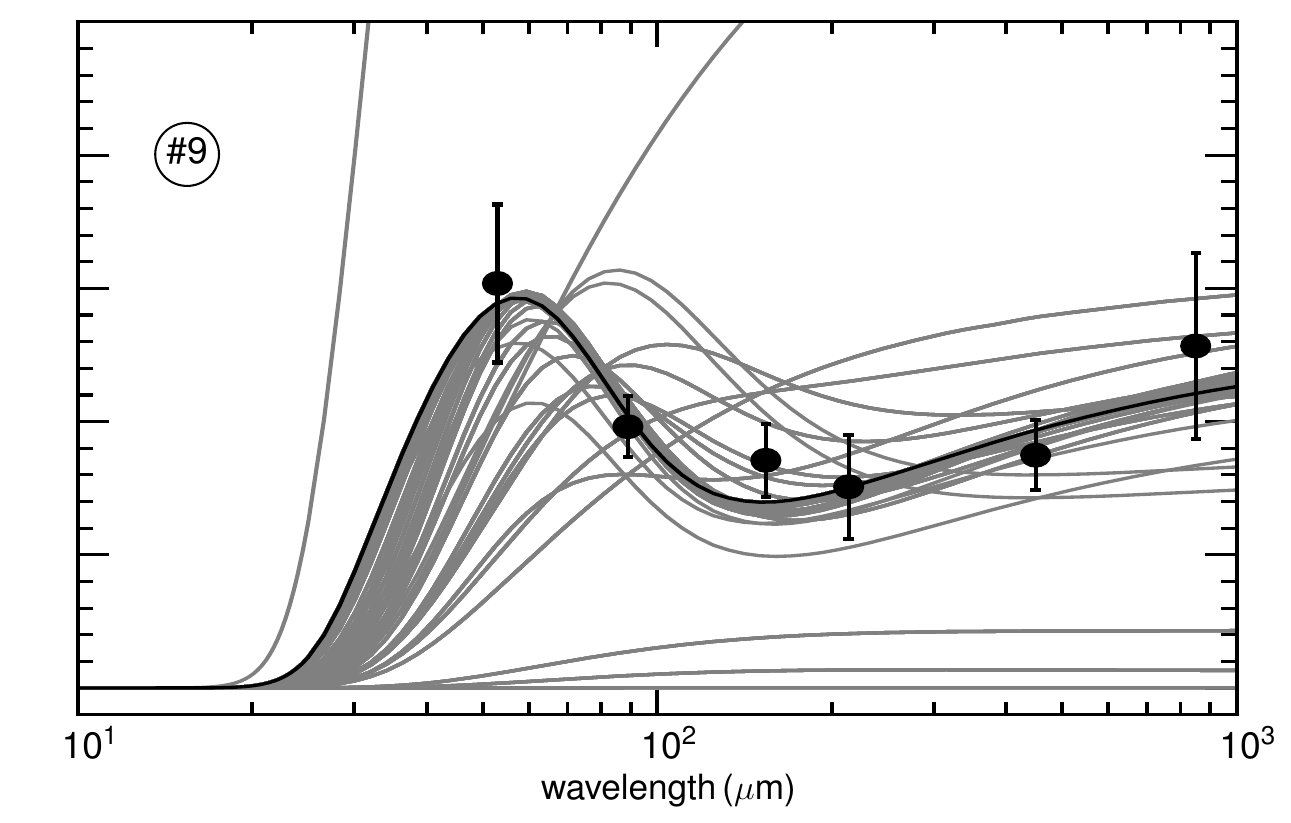}
    \caption{An example of the conversion of the \textsc{least-square} fitting procedure. The points are the data observed in \#9 as in Figure \ref{fig:modelvsobs_1phase}. The solid gray lines are the modelling predictions with parameter variations. The solid black line represents the best fit of the data. The fitting of the one-phase model is in the left panel, while the two-phase model is in the right panel.}
    \label{fig:curve_fit}
\end{figure*} 
As mentioned in Section \ref{sec:model}, we constrained the best parameters using the \textsc{least-square} function of the \textsc{LMFIT} library in Python. We set the boundaries of $f_{\rm max}$ in the intervals [0.01, 1.0], $\beta$ of the intervals [-5.0, -3.0] and $f_{\rm scale}$ of the intervals [1, $10^{4}$]. To speed up the fitting process, we manually adjust the intervals of $f_{\rm heat}$ for different regions. For example, $f_{\rm heat}$ is set lower than 3 for the majority, except for regions \#2, \#6 and \#7. Figure \ref{fig:curve_fit} shows an example of the conversion of the fit for the region \#9 with the one-phase model (left panel) and the two-phase model (right panel). The two-phase model significantly reproduces the observed spectrum and significantly improves the goodness of fit. It is worth mentioning that we experience that the best fit is quite sensitive to the free $f_{\rm scale}$ and $f_{\rm heat}$ boundaries (for example, the larger the boundary is set, the worse the fit is obtained). Thus, we narrowed these boundaries by running a serial test.
\end{document}